\documentclass[%
 superscriptaddress,
preprint,
 amsmath,amssymb,
 aps,
]{revtex4-2}

\usepackage{graphicx}
\usepackage{dcolumn}
\usepackage{bm}

\begin{document}

\preprint{APS/123-QED}

\title{Liquid Crystal Ground States on Cones with Anti-Twist Boundary Conditions}

\author{Cheng Long}
 \affiliation{%
 Department of Physics, Harvard University, Cambridge, Massachusetts 02138, USA
}%
\author{David R. Nelson}%
 \affiliation{%
 Department of Physics, Harvard University, Cambridge, Massachusetts 02138, USA
}%

\date{\today}

\begin{abstract}

Geometry and topology play a fundamental role in determining pattern formation on 2D surfaces in condensed matter physics. For example, local positive Gaussian curvature of a 2D surface attracts positive topological defects in a liquid crystal phase confined to the curved surface while repelling negative topological defects. Although the cone geometry is flat on the flanks, the concentrated Gaussian curvature at the cone apex geometrically frustrates liquid crystal orientational fields arbitrarily far away. The apex acts as an unquantized pseudo-defect interacting with the topological defects on the flank. By exploiting the conformal mapping methods of F. Vafa et al., we explore a simple theoretical framework to understand the ground states of liquid crystals with $p$-fold rotational symmetry on cones, and uncover important finite size effects for the ground states with boundary conditions that confine both plus and minus defects to the cone flanks. By combining the theory and simulations, we present new results for liquid crystal ground states on cones with anti-twist boundary conditions at the cone base, which enforce a total topological charge of $-1$. We find that additional quantized negative defects are created on the flank as the cone apex becomes sharper via a defect unbinding process, such that an equivalent number of quantized positive defects become trapped at the apex, thus partially screening the apex charge, whose magnitude is a continuous function of cone angle.

\end{abstract}

\maketitle


\section{Introduction}

Pattern formation on a two-dimensional curved surface in condensed matter physics is often associated with sophisticated, yet significant involvement of topology and geometry~\cite{bowick2009two}, from two-dimensional neutral superfluids~\cite{turner2010vortices,vitelli2004anomalous}, colloids confined at liquid interfaces~\cite{bausch2003grain,lipowsky2005direct,vitelli2006crystallography,bowick2007dynamics,irvine2010pleats,irvine2012fractionalization}, liquid crystals on curved surfaces with either planar or homeotropic surface anchoring~\cite{selinger2011monte,senyuk2013topological,koning2016crystals,nitschke2018nematic,koning2014saddle,pairam2013stable,lopez2011frustrated,koning2013bivalent}, to amphiphilic membranes~\cite{helfrich1985effect,nelson1987fluctuations,seung1988defects,baumgart2003imaging,mcmahon2005membrane,nelson2004statistical} and protein shells of viruses~\cite{ganser1999assembly,li2000image,lidmar2003virus,nguyen2005elasticity} in biological systems. A simple illustrative example for the role of topology in pattern formation of liquid crystals exploits the Poincar\'e-Hopf theorem: a spherical colloid immersed in a nematic liquid crystal solvent with strong planar anchoring on the surface requires the total topological charge of the surface defects to be $+2$ while a colloidal torus with planar anchoring in the same liquid crystal solvent requires the total topological charge of $0$. See, e.g. Ref~\cite{pairam2013stable,poulin1997novel}.

The liquid crystal orientational field on a curved surface depends not only on the topology of the surface, but also on its geometry. Although the topology of planar anchoring on a simple closed surface of genus $g=0$ enforces a global constraint on the total surface topological charge to be $+2$, the positions of the topological defects are actually determined by the surface geometry because the intrinsic geometry of the surface characterized by its Gaussian curvature creates a geometric potential field interacting with the topological defects so that positive topological defects gain lower energy moving to locations with larger positive Gaussian curvature and negative topological defects tend to move to locations with larger negative Gaussian curvature~\cite{tasinkevych2014dispersions}. The effects of surface geometry on the order parameter fields are crucial for understanding fundamental physics in two-dimensional condensed phases, as well as for illuminating many biological behaviors, like the role of curvature in the onset of spontaneous tissue flows in epithelial tissues~\cite{hannezo2011instabilities,shyer2013villification,callens2020substrate,harmand20213d,bell2022active} and the limb formation in regenerating \textit{Hydra}~\cite{maroudas2021topological}.

In contrast to curved surfaces whose Gaussian curvature is smoothed out, conical surfaces have Gaussian curvature concentrated at their apexes while the flanks are totally flat. The intrinsic geometric potential applied to a conical surface can be regarded as a potential field activated by an unquantized pseudo-defect placed at the apex whose topological charge depends on the total apex Gaussian curvature. This pseudo-defect caused by the apex Gaussian curvature triggers geometric frustration to the liquid crystal orientational fields on the conical surfaces, even with free boundary conditions at the cone base regardless of the cone size~\cite{zhang2022fractional}. On the other hand, the pseudo-defect charge also interacts with the topological defects on the flank, enabling the absorption of positive topological defects from the flank in the ground state as the conical apex becomes sharper~\cite{vafa2022defect}.

The goal of this article is to understand how positive Gaussian curvature concentrated at the apex of a conical surface affects liquid crystal ground states on the surface involving defects whose topological charges possess the opposite sign to the apex Gaussian curvature, which the previous theory did not explain properly~\cite{vafa2024defect}. In order to include negative topological defects on a cone with positive apex Gaussian curvature, we impose anti-twist boundary conditions at the cone base, as discussed below. In Section \MakeUppercase{\romannumeral 2}, we start from a generic distortion free energy of liquid crystals with $p$-fold rotational symmetry on a 2D curved surface caused by the intrinsic geometry. By employing the conformal mapping method for cones developed by F. Vafa et al., we demonstrate that the energetic ground states of liquid crystals on cones can be equivalently transformed into the ground states of 2D liquid crystals in a flat disk with an unquantized pseudo-defect fixed at the center. We then identify extra defect self-energy terms which increase logarithmically with the system size in the total free energy of a general defect configuration. This theory is tested for liquid crystal ground states on cones with both free boundary conditions and tangential boundary conditions. In Section \MakeUppercase{\romannumeral 3}, we apply our analytic theory to study ground states of $p$-atic liquid crystals on cones with anti-twist boundary conditions enforcing the total topological charge of $-1$. Our theory predicts that as the cone becomes sharper, at some specific cone apex angles, the apex creates a pair of new topological defects, leaving a plus defect trapped at the apex and the minus defect pushed out to the flank. We find that the total number of $-1/p$ defects on the flank thus changes as a function of the cone apex angle. The extra defect self-energy terms cause the fact that the transition points between configurations with different number of flank defects have an important dependence on the size of the cone. In addition, we make predictions for how the positions of the flank defects change as the sharpness of the cone evolves. In Section \MakeUppercase{\romannumeral 4}, a simulation model for $p$-atic liquid crystals on a cone is introduced based on the Maier-Saupe lattice model. To find liquid crystal ground states, we slowly anneal the discretized order parameter field in our lattice model from high temperature to zero temperature using a Langevin thermostat. We then compare the ground states found in our simulation model to our theoretical predictions, finding good agreement. In Section \MakeUppercase{\romannumeral 5}, we summarize our findings.

\section{Continuum Theory for Liquid Crystals on Cones}

In the continuum limit, the distortion free energy of a liquid crystal orientational field $\boldsymbol{u}(\sigma_1, \sigma_2)$ with $p$-fold rotational symmetry confined in an arbitrary 2D curved surface described by two independent internal parameters $(\sigma_1, \sigma_2)$ is given by~\cite{mackintosh1991orientational,park1996topological,vitelli2006nematic}
\begin{equation}
  F = \frac{1}{2}p^2 \tilde{J} \int d\sigma_1d\sigma_2 \sqrt{g} D_{\alpha} u^{\beta} D^{\alpha} u_{\beta}.
  \label{eqn:genericdistortionfe}
\end{equation}
Here, $\boldsymbol{u}$ is a unit vector, and $\tilde{J}$ is related to the interaction strength between nearest neighbors in our Maier-Saupe lattice model. For our analytic theory, $p$ and $\tilde{J}$ can be absorbed into a single elastic constant, but to make comparison between our analytic theory and our simulations later, we keep $p$ and $\tilde{J}$ as independent parameters in our theory. Here, $g$ is the determinant of the metric tensor, and $D_{\alpha}$ is the covariant derivative with respect to the metric of the surface. Duplicated indices in Eqn.~(\ref{eqn:genericdistortionfe}) are contracted using Einstein summation convention.

\begin{figure}[h]
\includegraphics[width=0.45\textwidth]{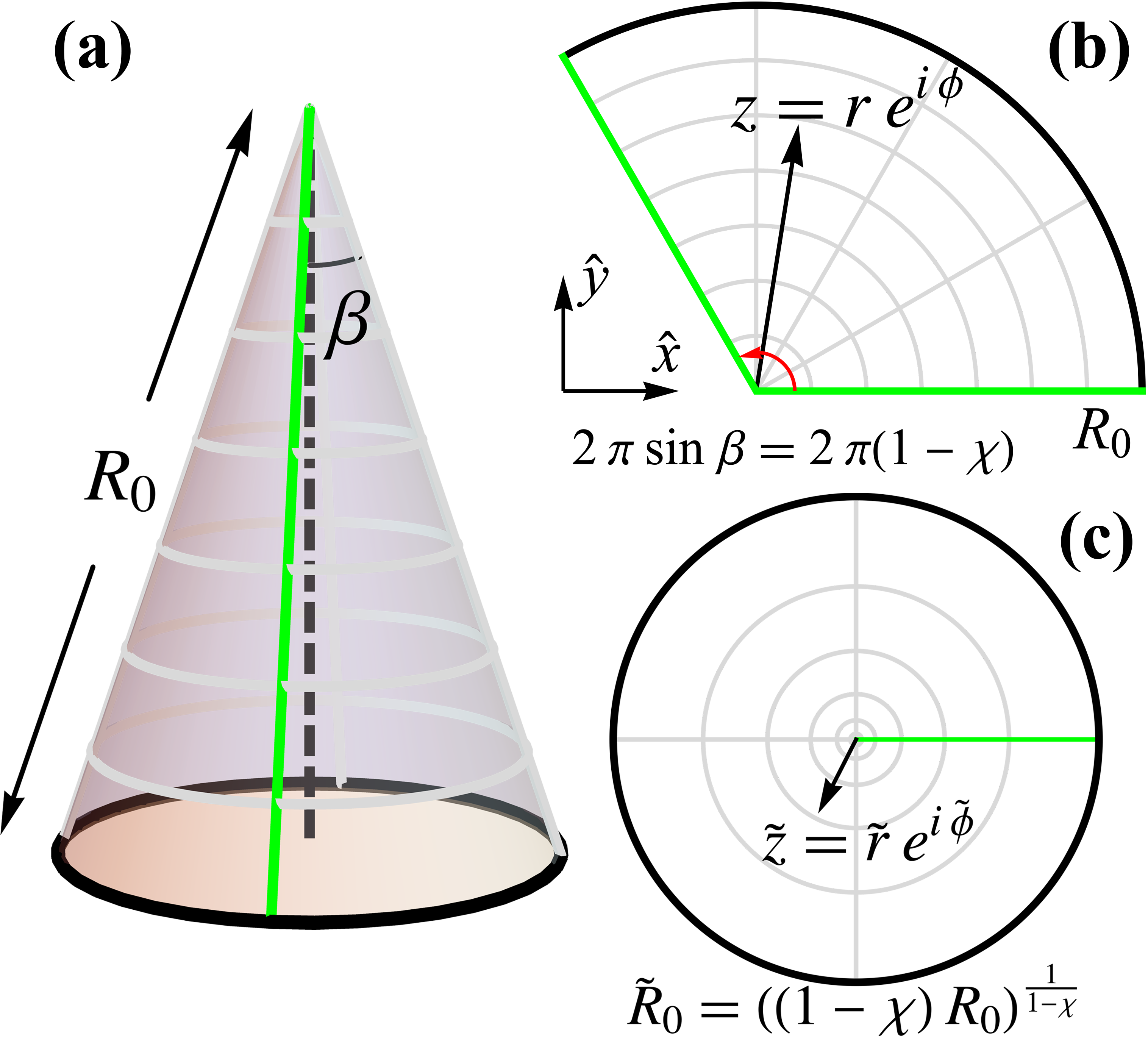}
\caption{\label{fig1:3dcone} (a) A 3D conical surface described by the apex half angle $\beta$ and the flank radius $R_0$. (b) The fundamental domain of the conical surface, constructed by cutting the cone along the green geodesic and unfolding it onto a flat plane. Each point in the fundamental domain can be described by a complex number using the polar coordinates, $z=re^{i\phi}$. The red sector angle is given by $2\pi \sin{\beta}=2\pi(1-\chi)$. (c) The conformal domain of the cone transformed from the fundamental domain by $\tilde{z}=\tilde{r}e^{i\tilde{\phi}}=\left((1-\chi)z\right)^{\frac{1}{1-\chi}}$. The evenly spaced grey circles on the original cone (a) are transformed into unevenly spaced circles in the conformal domain (c).}
\end{figure}
We now specialize to $p$-atic liquid crystals i.e. planar anisotropic liquids with a local $p$-fold rotational symmetry, where typically $p=1$, $2$, $4$ or $6$~\cite{vafa2024defect}, and these materials reside on a conical surface with strong planar anchoring, illustrated in Fig.~\ref{fig1:3dcone}(a). There are different choices for the internal coordinates $\sigma_1$ and $\sigma_2$ to describe the same conical metric space. Not all of them offer us a simple and physically intuitive Euler-Lagrange equation for $\boldsymbol{u}(\sigma_1,\sigma_2)$ to solve for ground states at zero temperature. In the previous work by G. Zhang and D. Nelson~\cite{zhang2022fractional}, cutting the 3D conical surface along a radial geodesic (the green line in Fig.~\ref{fig1:3dcone}(a)) and flattening it in a 2D Euclidean plane as shown in Fig.~\ref{fig1:3dcone}(b) leads to a free energy density exactly the same as the free energy density of 2D liquid crystals in a flat space,
\begin{equation}
  F = \frac{1}{2}p^2 \tilde{J} \int_0^{R_0} r dr \int_0^{2\pi(1-\chi)} d\phi \left| \boldsymbol{\nabla} \psi \right|^2.
  \label{eqn:fundamentaldomain}
\end{equation}
The flattened conical region is called the fundamental domain, which has the same radius $R_0$ as the conical flank radius. The sector angle between the two cut edges in the fundamental domain is determined by the apex half angle $\beta$ and is equal to $2\pi \sin{\beta}$. To simplify our description, we use the deficit angle parameter $\chi=1-\sin{\beta}$ to characterize the sharpness of the cone. When $\chi=0$, the conical surface is equivalent to a flat disk; when $\chi$ is greater than $0$, the surface is a pointy conical surface; as $\chi$ approaches $1$, the local geometry of the cone resembles the surface of a cylinder. In Eqn.~(\ref{eqn:fundamentaldomain}), $r$ and $\phi$ are planar polar coordinates, and the unit vector field $\boldsymbol{u}(r,\phi)$ is related to a scalar angle field $\psi(r,\phi)$ by $\boldsymbol{u}=\hat{\boldsymbol{x}}\cos{\psi}+\hat{\boldsymbol{y}}\sin{\psi}$, where $\hat{\boldsymbol{x}}$ and $\hat{\boldsymbol{y}}$ are the basis vectors of a 2D Cartesian coordinate system. Although the Euler-Lagrange equation for $\psi(r,\phi)$ simplifies in the fundamental domain, the parallel transport for an orientational field on a conical surface is manifested as a complicated boundary condition applied to the two cut edges in Fig.~\ref{fig1:3dcone}(b) to ensure the continuity of the orientational angle. The boundary condition reads~\cite{zhang2022fractional}
\begin{equation}
  \psi(r,0) = \psi(r,2\pi(1-\chi)) + 2\pi\chi + s\frac{2\pi}{p}.
  \label{eqn:continuityfundamentaldomain}
\end{equation}
On the right hand side, $2\pi \chi$ is the rotation or deficit angle obtained by parallel transporting a vector from point $(r,0)$ to $(r,2\pi(1-\chi))$ along the azimuthal direction. The $s/p$ term accounts for any topological charge generated by defects inside the arc defined by these two points.

To simplify the boundary condition in the fundamental domain, we resort to the conformal mapping method of F. Vafa et al.~\cite{vafa2022defect}, who exploited a conformal transformation that maps the fundamental domain of an unrolled cone to a full disk by a change of variables in the complex plane $\tilde{z}=\left((1-\chi)z\right)^{\frac{1}{1-\chi}}$. The conformal domain is illustrated in Fig.~\ref{fig1:3dcone}(c). In the conformal transformation, the complex variable $z=re^{i\phi}$ represents a point described by its polar coordinates in the fundamental domain, and $\tilde{z}=\tilde{r}e^{i\tilde{\phi}}$ is the associated point characterized by its polar coordinates in the conformal domain. The conformal transformation thus provides an explicit relation between the radial and angular coordinates in these domains,
\begin{equation}
  \begin{aligned}
    \tilde{r} &= \left((1-\chi)r\right)^{\frac{1}{1-\chi}} \\
    \tilde{\phi} &= \frac{\phi}{1-\chi}.
  \end{aligned}
  \label{eqn:conformaltrans}
\end{equation}
We can see from Eqn.~(\ref{eqn:conformaltrans}) that all the points in the fundamental domain are rotated counterclockwise due to the transformation from $\phi$ to $\tilde{\phi}$ so that the cut edge at $(r,2\pi(1-\chi))$ is pulled to overlap with the other cut edge in the conformal domain, and that the irregular shape of the fundamental domain becomes a full disk with a modified radius $\tilde{R}_0=\left((1-\chi)R_0\right)^{\frac{1}{1-\chi}}$ in the conformal domain. In addition to this angular stretching, the conformal transformation from $r$ to $\tilde{r}$ also changes the ratio of the radial coordinate to the radius of the domain, which leads to an uneven spacing between adjacent circles in the conformal domain transformed from equally spaced arcs in the fundamental domain, as illustrated in Fig.~\ref{fig1:3dcone}(b) and (c).

Upon applying the conformal transformation in Eqn.~(\ref{eqn:conformaltrans}) to the distortion free energy represented in the fundamental domain in Eqn.~(\ref{eqn:fundamentaldomain}), the free energy density is in fact invariant under the transformation, and the distortion free energy in the conformal domain reads
\begin{equation}
  F = \frac{1}{2}p^2 \tilde{J} \int_0^{\tilde{R}_0} \tilde{r} d\tilde{r} \int_0^{2\pi} d\tilde{\phi} \left| \boldsymbol{\nabla} \psi \right|^2.
  \label{eqn:conformaldomain}
\end{equation}
Even though it seems that the problem has been mapped to a simple liquid crystal problem in a 2D disk, we still have a discontinuous boundary condition imposed on the two overlapped cut edges in the conformal domain, which is now given by
\begin{equation}
  \psi(\tilde{r},0) = \psi(\tilde{r},2\pi) + 2\pi\chi + s\frac{2\pi}{p}.
  \label{eqn:discontinuousboundary}
\end{equation}
To address this discontinuous boundary condition, we subtract an unquantized pseudo-defect solution of $-\chi\tilde{\phi}$ from the scalar field $\psi$ and solve for another scalar field $\psi'=\psi+\chi\tilde{\phi}$ instead. In terms of the new scalar field $\psi'$, the boundary condition in Eqn.~(\ref{eqn:discontinuousboundary}) becomes
\begin{equation}
  \psi'(\tilde{r},0) = \psi'(\tilde{r},2\pi) + s\frac{2\pi}{p},
\end{equation}
and we can finally treat $\psi'(\tilde{r},\tilde{\phi})$ as a continuous field in the conformal domain without worrying about the boundary condition on the two overlapped cut edges, except for effects due to the topological defects on the cone flanks. The corresponding distortion free energy is written as
\begin{equation}
  F = \frac{1}{2}p^2 \tilde{J} \int_0^{\tilde{R}_0} \tilde{r} d\tilde{r} \int_0^{2\pi} d\tilde{\phi} \left| \boldsymbol{\nabla} \left(\psi'-\chi\tilde{\phi}\right) \right|^2.
  \label{eqn:feforpsiprime}
\end{equation}
Because $-\chi\tilde{\phi}$ is a solution to the Laplace equation with a pseudo-defect of charge $-\chi$ at the origin, the Euler-Lagrange equation for $\psi'(\tilde{r},\tilde{\phi})$ remains a Laplace equation, and solving for $\psi'(\tilde{r},\tilde{\phi})$ with given boundary conditions at the base of the conical surface now has a simple correspondence to a liquid crystal problem in a 2D disk, with an additional unquantized pseudo-defect fixed at the center. 

To understand Eqn.~(\ref{eqn:feforpsiprime}), note that the unquantized pseudo-defect at the origin comes from the concentrated Gaussian curvature at the apex of the conical surface, and this pseudo-defect creates a geometric potential on the cone interacting with the liquid crystal orientational fields confined on the surface. The pseudo-defect carries an unquantized charge of $-\chi$ dependent on the sharpness of the apex; this pseudo-defect can also interact with actual topological defects on the flank of the cone~\cite{vafa2022defect}, which are included in $\psi'(\tilde{r},\tilde{\phi})$. Since the pseudo-defect has negative charge for positive $\chi$, it attracts positive defects and repels negative defects on the flank, which agrees with how Gaussian curvature is expected to interact with 2D liquid crystals~\cite{vitelli2004anomalous}. As the cone becomes sharper, the interaction between the pseudo-defect and a topological defect on the flank also gets stronger. On the other hand, because of the pseudo-defect, the trivial solution of $\psi'(\tilde{r},\tilde{\phi})=\mathrm{const.}$ does not necessarily minimize the total free energy in Eqn.~(\ref{eqn:feforpsiprime}) for large $\chi$. A defect solution for $\psi'$ with topological charge opposite to the pseudo-defect can reduce the total distortion free energy by partially or fully screening out the pseudo-defect. 

Thus, to find the solutions of $\psi'(\tilde{r},\tilde{\phi})$ that minimize the total free energy, we consider a general solution of the Laplace equation including $N$ topological defects located on the flank of the cone, which is
\begin{equation}
  \psi'(\tilde{r},\tilde{\phi}) = \sum_{k=1}^{N} \left( \sigma_k D \left( \tilde{\boldsymbol{r}}; \tilde{\boldsymbol{r}}_k \right) + \sigma_k^{\mathrm{Im}} D \left( \tilde{\boldsymbol{r}}; \tilde{\boldsymbol{r}}_k^{\mathrm{Im}} \right)  \right)
  \label{eqn:generalsol}
\end{equation}
where $\sigma_k D \left( \tilde{\boldsymbol{r}}; \tilde{\boldsymbol{r}}_k \right)$ is the solution for a single defect of charge $\sigma_k$ located at $\tilde{\boldsymbol{r}}_k$ inside the conformal domain,
\begin{equation}
  \sigma_k D \left( \tilde{\boldsymbol{r}}; \tilde{\boldsymbol{r}}_k \right) = \sigma_k \tan^{-1}{\left(\frac{\tilde{y}-\tilde{y}_k}{\tilde{x}-\tilde{x}_k}\right)}.
  \label{eqn:singledefectsol}
\end{equation}
In Eqn.~(\ref{eqn:singledefectsol}), $(\tilde{x},\tilde{y})$ is a set of Cartesian coordinates in the conformal domain, which can be easily related to the conformal polar coordinates by $(\tilde{r}\cos{\tilde{\phi}},\tilde{r}\sin{\tilde{\phi}})$. For the flank defects considered in Eqn.~(\ref{eqn:generalsol}), the range of $\lvert \tilde{\boldsymbol{r}}_k \rvert$ is $0<\lvert \tilde{\boldsymbol{r}}_k \rvert < \tilde{R}_0$. To account for absorbed topological defects of total charge $\sigma_0$ at the apex, we combine those defects with the pseudo-defect term in Eqn.~(\ref{eqn:feforpsiprime}) and change the apex charge from $-\chi$ to $-\chi+\sigma_0$. Although linear combination of single defect solutions satisfies the Euler-Lagrange equation for $\psi'$, we still have to consider how to satisfy the boundary conditions applied to the base of the cone. Image defect contributions, $\sigma_k^{\mathrm{Im}} D \left( \tilde{\boldsymbol{r}}; \tilde{\boldsymbol{r}}_k^{\mathrm{Im}} \right)$, are therefore added to Eqn.~(\ref{eqn:generalsol}) in order to address the boundary conditions. The positions of the image defects are placed outside the conformal domain, and both $\sigma_k^{\mathrm{Im}}$ and $\tilde{\boldsymbol{r}}_k^{\mathrm{Im}}$ are dependent on the specific boundary conditions at the base of the cone, as discussed below. For our analytical theory, we discuss three types of common boundary conditions at the cone base, which are free boundary conditions, tangential boundary conditions, and anti-twist boundary conditions. Free and tangential boundary conditions were considered in Ref.~\cite{zhang2022fractional} and~\cite{vafa2022defect}, but we discuss them here for completeness and comparison. 

Regarding free boundary conditions, there is no preferred orientation for the scalar order parameter $\psi'(\tilde{R}_0, \tilde{\phi})$ at the boundary, but $\psi'(\tilde{R}_0,\tilde{\phi})$ has to satisfy a mechanical equilibrium condition instead. It is easy to show that to satisfy the free boundary conditions at the base of a cone, for each defect of charge $\sigma_k$ at $\tilde{z}_k=\tilde{r}_k e^{i\tilde{\phi}_k}$ inside the conformal domain, a corresponding image defect of charge $-\sigma_k$ has to be added at $\tilde{z}_k^{\mathrm{Im}}=\tilde{R}_0^2/\bar{\tilde{z}}_k$. The overbar on $\bar{\tilde{z}}_k$ means the complex conjugate of $\tilde{z}_k$. Since the relevant image defects carry the opposite charge of the real defects, the free boundary conditions attract those real defects toward the boundary and can readily annihilate them.

\begin{figure}[h]
\includegraphics[width=0.45\textwidth]{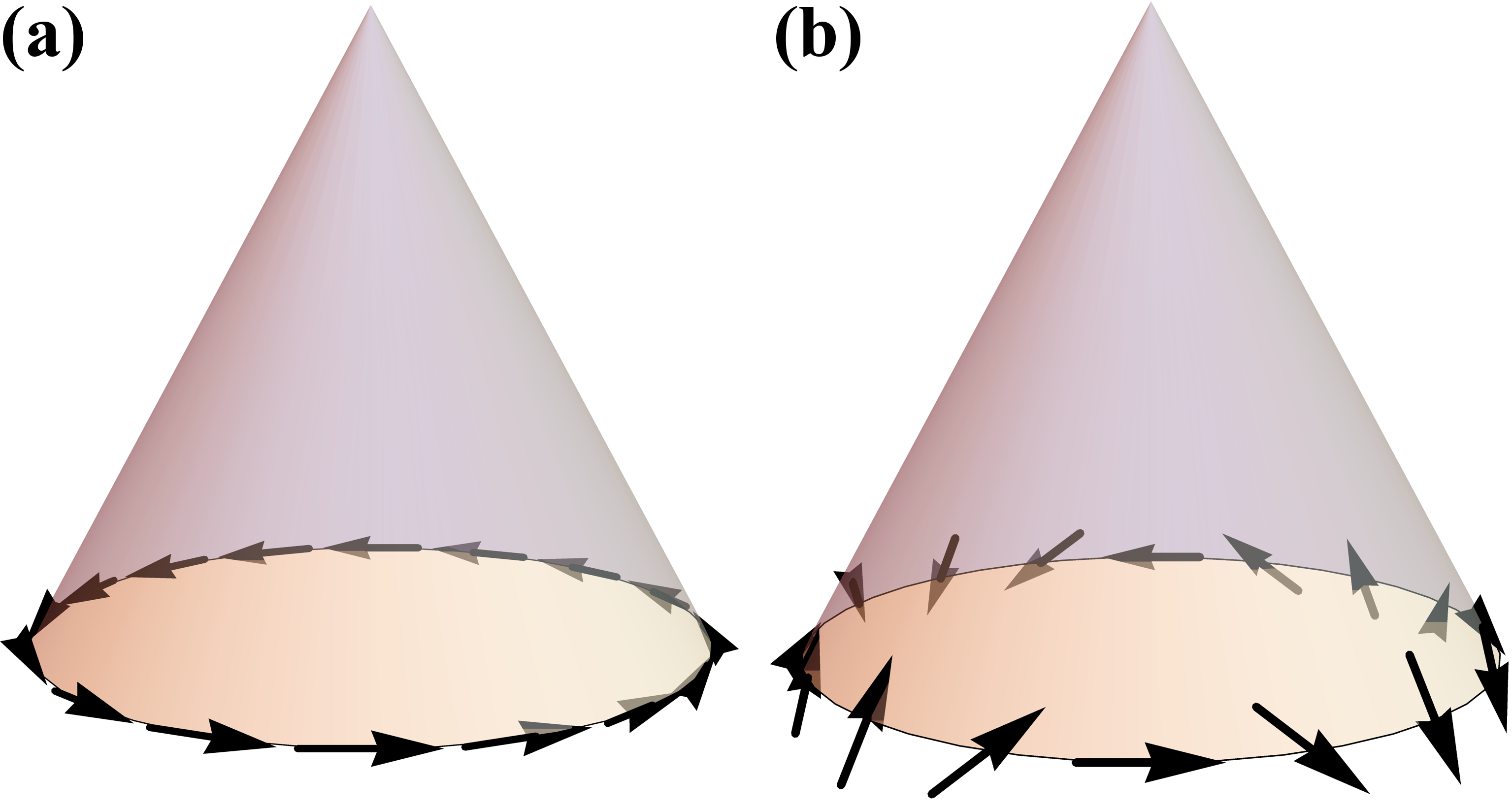}
\caption{\label{fig2:boundaryconditions} (a) Tangential boundary conditions at the base of a cone. (b) Anti-twist boundary conditions at the base of a cone. Black arrows represent the orientations of the order parameter at the boundary.}
\end{figure}
Tangential boundary conditions (Fig.~\ref{fig2:boundaryconditions}(a)), on the other hand, require the orientational order parameter at the base of a cone to be always parallel to the tangential direction of the circular boundary, so if measured in a fixed laboratory frame whose $z$-axis is aligned with the conical axis, the order parameter is rotated by $2\pi$ around the $z$-axis, moving a full circle around the base. If we choose instead to measure the orientation of the order parameter in a locally varying orthonormal basis corresponding to the flank radial direction away from the apex and the azimuthal direction, the rotation angle of the order parameter is $0$ because the order parameter rotates synchronously with the local azimuthal direction. The topology of tangential boundary conditions at the base of a cone thus enforces the total topological charge of $+1$ on the cone. 

Anti-twist boundary conditions (Fig.~\ref{fig2:boundaryconditions}(b)) also impose a preferred orientation for the order parameter at the boundary. In contrast to tangential boundary conditions, moving a full circle around the base, the orientation of the projection of the order parameter onto the $x$-$y$ plane is now rotated by $-2\pi$ around the $z$-axis in the fixed laboratory frame. Note that the order parameter for anti-twist boundary conditions does not stay in the same plane. In this case, the variation of the order parameter is more conveniently described in the locally varying coordinate basis mentioned above, and the rotation angle is then $-4\pi$. The topology of anti-twist boundary conditions thus gives rise to a constraint of the total topological charge being $-1$ on the cone. 

For either tangential boundary conditions or anti-twist boundary conditions, a real defect of charge $\sigma_k$ at $\tilde{z}_k=\tilde{r}_k e^{i\tilde{\phi}_k}$ inside the conformal domain has to be paired with an image defect of the same charge $\sigma_k$ at $\tilde{z}_k^{\mathrm{Im}}=\tilde{R}_0^2/\bar{\tilde{z}}_k$, and that indicates both tangential boundary conditions and anti-twist boundary conditions impose a repulsion to flank defects away from the boundary.

With the addition of image defects caused by the boundary conditions mentioned above, after inserting the solution of $\psi'$ in Eqn.~(\ref{eqn:generalsol}) into Eqn.~(\ref{eqn:feforpsiprime}), all the integrals inside the total free energy can be done analytically, and the total free energy of liquid crystals on a cone can be written explicitly as a function of the defect charges $\sigma_k$ and the defect positions $\tilde{\boldsymbol{r}}_k$. Here, we derive the total free energy for a general $N$-defect configuration with anti-twist boundary conditions. The total free energy for tangential boundary conditions actually shares the same form as that for anti-twist boundary conditions, since the image defects for both cases have the same charges and positions. It is the constraint on the total topological charge on the cone ($+1$ or $-1$ for tangential and anti-twist boundary conditions, respectively) that sets the difference between these two boundary condition choices. Anti-twist boundary conditions on the cone are important because they allow us to study topological defects with negative charges in the ground state, in contrast to the previous studies~\cite{zhang2022fractional,vafa2022defect}.

As noted above, for anti-twist boundary conditions, we have $\sigma_k^{\mathrm{Im}}=\sigma_k$ and $\tilde{z}_k^{\mathrm{Im}}=\tilde{R}_0^2/\bar{\tilde{z}}_k$. Upon noticing that the pseudo-defect term from the apex is in fact equal to $-\chi D\left(\tilde{\boldsymbol{r}};0\right)$, calculating the integrals in Eqn.~(\ref{eqn:feforpsiprime}) amounts to evaluating the integral of the following integrand over the entire conformal domain,
\begin{equation}
\boldsymbol{\nabla}\! \left(D\left( \tilde{\boldsymbol{r}};\tilde{\boldsymbol{r}}_m \right) + D\left( \tilde{\boldsymbol{r}};\tilde{\boldsymbol{r}}_m^{\mathrm{Im}} \right) \right)\cdot \boldsymbol{\nabla}\! \left(D\left( \tilde{\boldsymbol{r}};\tilde{\boldsymbol{r}}_n \right) + D\left( \tilde{\boldsymbol{r}};\tilde{\boldsymbol{r}}_n^{\mathrm{Im}} \right) \right).
\end{equation}
After multiplying out the integrand above, one can do the integration term by term. To make the entire derivation concise, we supply the Green's function for a defect of charge $+1$ at $\tilde{\boldsymbol{r}}_m$ paired with its like-signed image defect in the conformal domain. The Green's function reads as in Ref~\cite{vafa2022defect}
\begin{equation}
  G\left(\tilde{z};\tilde{z}_m\right) = \frac{1}{2}\ln{\frac{\lvert \tilde{z} - \tilde{z}_m \rvert^2}{\tilde{R}_0^2}} + \frac{1}{2}\ln{\lvert 1 - \frac{\tilde{z}\bar{\tilde{z}}_m}{\tilde{R}_0^2} \rvert^2},
\end{equation}
where the overbar again indicates complex conjugation. The Laplacian of the Green's function produces the charge distribution of an equivalent 2D Coulomb system:
\begin{equation}
  \nabla^2 G\left(\tilde{z};\tilde{z}_m\right) = 2\pi \left( \delta\left( \tilde{z}-\tilde{z}_m \right) + \delta\left( \tilde{z}-\frac{\tilde{R}_0^2}{\bar{\tilde{z}}_m} \right) \right).
  \label{eqn:laplaciangreenfunc}
\end{equation}
Here, $ \delta\left( \tilde{z}-\tilde{z}_m \right)$ is a Dirac delta function with the singularity at $\tilde{z}_m$. The first Dirac-delta term in the charge distribution in Eqn.~(\ref{eqn:laplaciangreenfunc}) corresponds to the defect at $\tilde{z}_m$; the second Dirac-delta term corresponds to its image defect at $\tilde{R}_0^2/\bar{\tilde{z}}_m$. For free boundary conditions where $\sigma_k^{\mathrm{Im}}=-\sigma_k$, the sign before the second term in the Green's function changes from plus to minus. With the help of this Green's function, the integral of the defect cross term between $\tilde{\boldsymbol{r}}_m$ and $\tilde{\boldsymbol{r}}_n$ ($m \neq n$) normalized by the product of their topological charges becomes
\begin{equation}
  \begin{aligned}
    &\int\! d\tilde{\boldsymbol{r}} \boldsymbol{\nabla}\! \left(D\left( \tilde{\boldsymbol{r}};\tilde{\boldsymbol{r}}_m \right)\! +\! D\left( \tilde{\boldsymbol{r}};\tilde{\boldsymbol{r}}_m^{\mathrm{Im}} \right) \right)\! \cdot \! \boldsymbol{\nabla}\! \left(D\left( \tilde{\boldsymbol{r}};\tilde{\boldsymbol{r}}_n \right)\! +\! D\left( \tilde{\boldsymbol{r}};\tilde{\boldsymbol{r}}_n^{\mathrm{Im}} \right) \right) \\
    &= \int d\tilde{\boldsymbol{r}}\: \boldsymbol{\nabla} G\left( \tilde{z};\tilde{z}_m \right) \cdot \boldsymbol{\nabla} G\left( \tilde{z};\tilde{z}_n \right) \\
    &=-2\pi G\left( \tilde{z}_m;\tilde{z}_n \right).
  \end{aligned}
  \label{eqn:crossinteracting}
\end{equation}
The surface term coming from integration by parts in the Green's function vanishes provided the defects at $\tilde{\boldsymbol{r}}_m$ and $\tilde{\boldsymbol{r}}_n$ are located inside the conformal domain. The cross terms describe the interaction between flank defects and their corresponding image charges.

In addition to the cross terms between different defects ($m \neq n$), the total free energy also includes self-interacting terms ($m = n$) which can also be calculated in a similar manner. A similar Green's function analysis tells us in this case that
\begin{equation}
  \begin{aligned}
    &\int d\tilde{\boldsymbol{r}}\: \lvert \boldsymbol{\nabla}\! \left(D\left( \tilde{\boldsymbol{r}};\tilde{\boldsymbol{r}}_k \right) + D\left( \tilde{\boldsymbol{r}};\tilde{\boldsymbol{r}}_k^{\mathrm{Im}} \right) \right) \rvert^2 \\
    &= -2\pi G\left( \tilde{z}_k;\tilde{z}_k \right),
  \end{aligned}
\end{equation}
which clearly diverges as $\lim_{\tilde{z}\to \tilde{z}_k}{\lvert \tilde{z}-\tilde{z}_k \rvert^2}$. This divergence is expected because in 2D nematic liquid crystals, the total free energy of a director field including a topological defect diverges as the defect core size goes to zero, and we have to define a finite defect core size below which the magnitude of the orientational order parameter is driven to zero due to the large spatial variation of the director. Here, we introduce $\tilde{\delta}_k$ as the defect core radius for a defect at $\tilde{z}_k$ in the conformal domain. Since the Maier-Saupe simulations in Section~\ref{sec:simulations} are carried out on a lattice in the fundamental domain where the defect core radius can be defined as half the lattice spacing $a/2$, according to the conformal transformation in Eqn.~(\ref{eqn:conformaltrans}), \textit{the defect core radius in the conformal domain must depend on the defect position}. Turner and Vitelli argued in a similar fashion in Ref~\cite{vitelli2004anomalous}. On assuming the defect is located at a distance much greater than the defect core radius away from the apex, the relation between its core radius in the fundamental domain and that in the conformal domain is in fact~\cite{turner2010vortices,vitelli2004anomalous} $\tilde{\delta}_k = (a/2)\tilde{r}_k^{\chi}$. Upon adding this ultraviolet cutoff to the Green's function $G\left( \tilde{z}_k,\tilde{z}_k \right)$, the self-interacting terms in the total free energy become
\begin{equation}
  \begin{aligned}
    &\int d\tilde{\boldsymbol{r}}\: \lvert \boldsymbol{\nabla}\! \left(D\left( \tilde{\boldsymbol{r}};\tilde{\boldsymbol{r}}_k \right) + D\left( \tilde{\boldsymbol{r}};\tilde{\boldsymbol{r}}_k^{\mathrm{Im}} \right) \right) \rvert^2 \\
    &= -2\pi \left( \frac{\chi}{2}\ln{\frac{\lvert \tilde{z}_k \rvert^2}{\tilde{R}_0^2}} - \ln{\frac{(1-\chi)R_0}{a}} + \ln{\left(1-\frac{\lvert \tilde{z}_k \rvert^2}{\tilde{R}_0^2}\right)} \right).
  \end{aligned}
  \label{eqn:greensfunctionself}
\end{equation}

Finally, after combining the cross terms and the self-interacting terms for all the topological defects, as well as the contribution from the unquantized pseudo-defect at the apex combined with possible absorbed topological defects of total charge $\sigma_0$, the total free energy of the $N$-defect configuration on a cone with either anti-twist boundary conditions or tangential boundary conditions normalized by $\pi p^2 \tilde{J}$ becomes
\begin{equation}
  \begin{aligned}
    \frac{F}{\pi p^2 \tilde{J}} = - \sum_{m<n} \sigma_m \sigma_n \left( \ln{\frac{\lvert \tilde{z}_m - \tilde{z}_n \rvert^2}{\tilde{R}_0^2}} + \ln{\lvert 1 - \frac{\tilde{z}_m\bar{\tilde{z}}_n}{\tilde{R}_0^2} \rvert^2} \right) \\
    - \left(\sigma_0 - \chi \right) \sum_{k=1}^{N}{\sigma_k \ln{\frac{\lvert \tilde{z}_k \rvert^2}{\tilde{R}_0^2}}} - \sum_{k=1}^{N} \sigma_k^2 \ln{\left( 1- \frac{\lvert \tilde{z}_k \rvert^2}{\tilde{R}_0^2} \right)} \\
    - \chi \sum_{k=1}^{N} \frac{\sigma_k^2}{2} \ln{\frac{\lvert \tilde{z}_k \rvert^2}{\tilde{R}_0^2}} \! +\! \sum_{k=1}^{N}\! \sigma_k^2 \ln{\frac{(1\!-\! \chi)R_0}{a}}\! +\! \frac{(\sigma_0 \!-\! \chi)^2}{1\!-\! \chi}\! \ln\!{\frac{R_0}{a}}.
  \end{aligned}
  \label{eqn:totalfreeenergyndefect}
\end{equation}
Note that the total free energy of liquid crystals on a cone is now expressed as an algebraic function of defect charges and defect positions in the conformal domain, with boundary conditions encoded by the position and topological charge of the image defects. As discussed above, for anti-twist boundary conditions, we have an additional constraint on the total topological charge, $\sum_{k=0}^N{\sigma_k}=-1$. For tangential boundary conditions, we have $\sum_{k=0}^N{\sigma_k}=+1$~\cite{vafa2022defect}.

The first term in Eqn.~(\ref{eqn:totalfreeenergyndefect}) describes the interaction between different topological defects on the flank and with their image defects. Note that the magnitude of this term is proportional to the product of the two different defect charges, and it is derived directly from the Green's function in Eqn.~(\ref{eqn:crossinteracting}). Inside the parenthesis of the first term, we have the combined energy from the interaction between $\sigma_m$ and $\sigma_n$ and that between $\sigma_m^{\mathrm{Im}}$ and $\sigma_n^{\mathrm{Im}}$, embodied in the first part of the parenthesis. As a consequence, when those two defects carry charges of the same sign, the first part predicts a repulsive force between the two defects with a strength proportional to $\lvert \tilde{z}_m-\tilde{z}_n \rvert^{-1}$. The remaining term in the parenthesis arises from the interaction between the defects and the boundary conditions, i.e. from the interaction between $\sigma_m$ and $\sigma_n^{\mathrm{Im}}$ and that between $\sigma_m^{\mathrm{Im}}$ and $\sigma_n$, so the sign in front of this part depends on the type of boundary conditions. For either anti-twist or tangential boundary conditions, the image defect of a topological defect carries the same charge as the actual defect, so this part adds a force pushing either $\sigma_m$ or $\sigma_n$ towards the apex. The second term in the total free energy comes from the interaction between the defects on the flank and the combination of the pseudo-defect and the absorbed defects at the apex so this energy term is proportional to $(\sigma_0-\chi)\sigma_k$. This second term gives rise to a force between the charges $\sigma_0-\chi$ and $\sigma_k$, again inversely proportional to the distance between them.

The remaining terms in Eqn.~(\ref{eqn:totalfreeenergyndefect}) are derived from the self-interacting term of each defect (including the pseudo-defect at the apex) using the result of Eqn.~(\ref{eqn:greensfunctionself}). The third, the fourth, and the fifth terms arise from the self-interacting terms of the $N$ defects on the flank combined with their image defects, and all of them are proportional to $\sigma_k^2$. Physically, the third term characterizes the remaining part of the interaction between the $N$ defects on the flank and the boundary conditions at the base of the cone. For both tangential and anti-twist boundary conditions, it implies a direct repulsion between the defect and the boundary. For free boundary conditions, the sign of this term would be positive, implying an attraction between the defect and the boundary. The fourth term in the total free energy is caused by the dependence of the defect core size on the geometry of the 2D surface, which was first predicted by V. Vitelli and A. Turner~\cite{vitelli2004anomalous}. For a defect on a conical surface, in addition to the interaction directly caused by the pseudo-defect, the positive Gaussian curvature concentrated at the apex pushes the defect away from the apex regardless of the sign of the defect charge. This term thus leads to an attractive force between the apex and a $+1/p$ defect that is weaker than the repulsive force between the apex and a $-1/p$ defect at the same location. The fifth term accounts for the generalization of the self-energy of each defect on the flank of a conical surface. When $\chi=0$, this term contributes to a sum over $\sigma_k^2 \ln{(R_0/a)}$, a well-known result for the self-energy of a defect with charge $\sigma_k$ on a flat surface. The factor $1-\chi$ inside the logarithm appears more generally due to the reduced area of the fundamental domain for a cone. In the thermodynamic limit of $R_0 \to \infty$, the $1-\chi$ factor is subleading compared to $R_0/a$ inside the logarithm. When $R_0/a$ is finite, the $1-\chi$ factor does play a role in the total free energy, and it is detectable from our simulation data, which is demonstrated in the Appendix~\ref{appendixa}. The $1-\chi$ factor is also a reminder that as $\chi \to 1$, the self-energy on a finite cone diverges, because the radius of the cone base shrinks as $(1-\chi)R_0$. We note that in order to consider a very sharp cone ($\chi \to 1$) with a finite radius at the cone base compared to the defect core size, the flank radius has to be always greater than $a(1-\chi)^{-1}$.

The last term in Eqn.~(\ref{eqn:totalfreeenergyndefect}) represents the self-energy of the pseudo-defect combined with possible absorbed defects at the apex, which is therefore proportional to $(\sigma_0-\chi)^2$. To make a clearer comparison between our theory and simulations, we assume the defect core radius for the apex is equal to $a$ in the fundamental domain, slightly bigger than $a/2$ for those quantized defects on the flanks. 

In the next section, we will show that the last two terms in Eqn.~(\ref{eqn:totalfreeenergyndefect}), although independent of the positions of defects on the flank, play an important role in determining ground state energies and the transition boundaries between different ground states for liquid crystals on cones with anti-twist boundary conditions, as a function of the deficit angle $2\pi \chi$ and the system size $R_0$.

In a similar manner, we can also derive the total free energy of a $N$-defect configuration on a cone with free boundary conditions where $\sigma_k^{\mathrm{Im}}=-\sigma_k$, the situation studied in Ref.~\cite{zhang2022fractional}. Eqn.~(\ref{eqn:totalfreeenergyndefect}) is now replaced by
\begin{equation}
  \begin{aligned}
    \frac{F}{\pi p^2 \tilde{J}} = - \sum_{m<n} \sigma_m \sigma_n \left( \ln{\frac{\lvert \tilde{z}_m - \tilde{z}_n \rvert^2}{\tilde{R}_0^2}} - \ln{\lvert 1 - \frac{\tilde{z}_m\bar{\tilde{z}}_n}{\tilde{R}_0^2} \rvert^2} \right) \\
    - \left(\sigma_0 - \chi \right) \sum_{k=1}^{N}{\sigma_k \ln{\frac{\lvert \tilde{z}_k \rvert^2}{\tilde{R}_0^2}}} + \sum_{k=1}^{N} \sigma_k^2 \ln{\left( 1- \frac{\lvert \tilde{z}_k \rvert^2}{\tilde{R}_0^2} \right)} \\
    - \chi \sum_{k=1}^{N} \frac{\sigma_k^2}{2} \ln{\frac{\lvert \tilde{z}_k \rvert^2}{\tilde{R}_0^2}} \! +\! \sum_{k=1}^{N}\! \sigma_k^2 \ln{\frac{(1\!-\! \chi)R_0}{a}}\! +\! \frac{(\sigma_0 \!-\! \chi)^2}{1\!-\! \chi}\! \ln\!{\frac{R_0}{a}}.
  \end{aligned}
  \label{eqn:totalfreeenergyndefectfreeboundary}
\end{equation}
Note the sign differences compared to Eqn.~(\ref{eqn:totalfreeenergyndefect}) in the final terms of the first and second lines. The physical meaning of each term in the total free energy above is similar to that in Eqn.~(\ref{eqn:totalfreeenergyndefect}) except that the free boundary conditions manifested in the final term of the second line attract topological defects towards the boundary instead of repelling them.

\subsection{Cones with Free Boundary Conditions}

A simple illustration of our analytic theory arises from using the total free energy in Eqn.~(\ref{eqn:totalfreeenergyndefectfreeboundary}) to determine the ground states of liquid crystals on cones with free boundary conditions. Free boundary conditions do not impose any constraints on the total topological charge for defects on the conical surface, and it is easy to see that the ground states should be configurations where the pseudo-defect of charge $-\chi$ at the apex is maximally neutralized by absorption of quantized topological defects with net charge $\sigma_0$. There are no flank defect in the ground states, and the total free energy in Eqn.~(\ref{eqn:totalfreeenergyndefectfreeboundary}) simplifies to
\begin{equation}
  \begin{aligned}
    \frac{F}{\pi p^2 \tilde{J}} =  \frac{(\sigma_0 - \chi)^2}{1- \chi} \ln{\frac{R_0}{a}},
  \end{aligned}
  \label{eqn:totalfreeenergyndefectfreeboundarysimp}
\end{equation}
which represents the self-energy of the residual charge at the apex. The ideal exact cancellation of the apex charge $-\chi$ by $\sigma_0$ would produce a total free energy minimized to zero, but the absorbed charge can only take values of $1/p$ multiplied by an integer. Accordingly, for a general conical surface with a well-defined $\chi$, the liquid crystal ground states are configurations with a topological defect of quantized charge $\sigma_0$ at the apex, whose value minimizes $\lvert \sigma_0 - \chi \rvert$. This result agrees with the free boundary condition results of Ref.~\cite{zhang2022fractional}.

\subsection{Cones with Tangential Boundary Conditions}

Although $p$-atic liquid crystals on cones with tangential boundary conditions have been studied previously~\cite{vafa2022defect}, the theory developed here includes finite size effects, which were not considered in the previous theory. We do recover the results of Ref.~\cite{vafa2022defect} for defect absorption transitions in the limiting case when the system size $R_0$ is infinitely large. More generally, we find that transitions between ground states with different number of flank defects depend on the ratio of the system size $R_0$ to the defect core size $a$. As we shall see, such effects are particularly important when we move on to treat anti-twist boundary conditions.

To find the ground states, we consider configurations in which there are a composite defect of charge $-e/p$ absorbed at the apex and $p+e$ flank defects of charge $+1/p$ distributed evenly around the apex at $\tilde{z}_k = \tilde{r}_0 e^{i 2\pi k/(p+e)}$. Here, $k=1,2,\dotsc,p+e$, and $e$ can be any integer equal or greater than $-p$ in principle. When $e$ is a positive integer, it means the number of extra $+1/p$ defects created on the flank. When $e$ is a negative integer, it means the number of $+1/p$ defects absorbed to the apex. With this assumption, the total topological charge on the cone is always fixed at $+1$, thus satisfying the tangential boundary conditions. Upon inserting the defect positions into the general free energy for a $N$-defect configuration in Eqn.~(\ref{eqn:totalfreeenergyndefect}), we get
\begin{equation}
  \begin{aligned}
    \frac{F}{\pi p^2 \tilde{J}} = -\frac{p+e}{p^2}\bigg\{ \ln{\left( (p+e) \bar{r}_0^{p+e-1} (1-\bar{r}_0^{2(p+e)}) \right)} + \\ 
    p\left( \frac{\chi}{2p} \!-\! \chi \!-\! \frac{e}{p} \right) \ln{\bar{r}_0^2}\! \bigg\} + \frac{p \!+\! e}{p^2} \ln{\frac{1 \!-\! \chi}{\bar{a}}} + \frac{(-\! \chi \!-\! e/p)^2}{1-\chi}\ln{\frac{1}{\bar{a}}}
  \end{aligned}
  \label{eqn:totalfreeenergytangential}
\end{equation}
where $\bar{r}_0 \equiv \tilde{r}_0/\tilde{R}_0$ and $\bar{a}\equiv a/R_0$. The last two terms dependent on the ratio $\bar{a}$ come from the self-energy of the defects and the apex. This free energy looks complicated, yet its physical interpretation is simple. By fixing the total number of flank defects $p+e$ and minimizing the total free energy with respect to $\bar{r}_0$, we find
\begin{equation}
  \bar{r}_0 = \left(\frac{p-e-1+\chi-2p\chi}{3p+e-1+\chi-2p\chi}\right)^{\frac{1}{2(p+e)}},
  \label{eqn:defectpostangential}
\end{equation}
thus reproducing the result of Ref~\cite{vafa2022defect} for tangential boundary conditions. The optimized position of defects with this ansatz is independent of the ratio $\bar{a}$. We then insert the optimized $\bar{r}_0$ into Eqn.~(\ref{eqn:totalfreeenergytangential}), and the optimum number of absorbed flank defects $e$ for the free energy minimization can be determined numerically as a function of $\chi$ for any predetermined $\bar{a}$. When $\chi=0$, which corresponds to a flat surface, the energetic ground state includes $p$ defects of charge $+1/p$ distributed evenly around the center of the disk. As the cone becomes sharper, the pseudo-defect charge $-\chi$ gets larger in magnitude, and absorbing plus defects to the apex becomes energetically favorable for cancelling the distortion energy caused by the pseudo-defect at the apex. 

\begin{figure*}[h]
\includegraphics[width=0.60\textwidth]{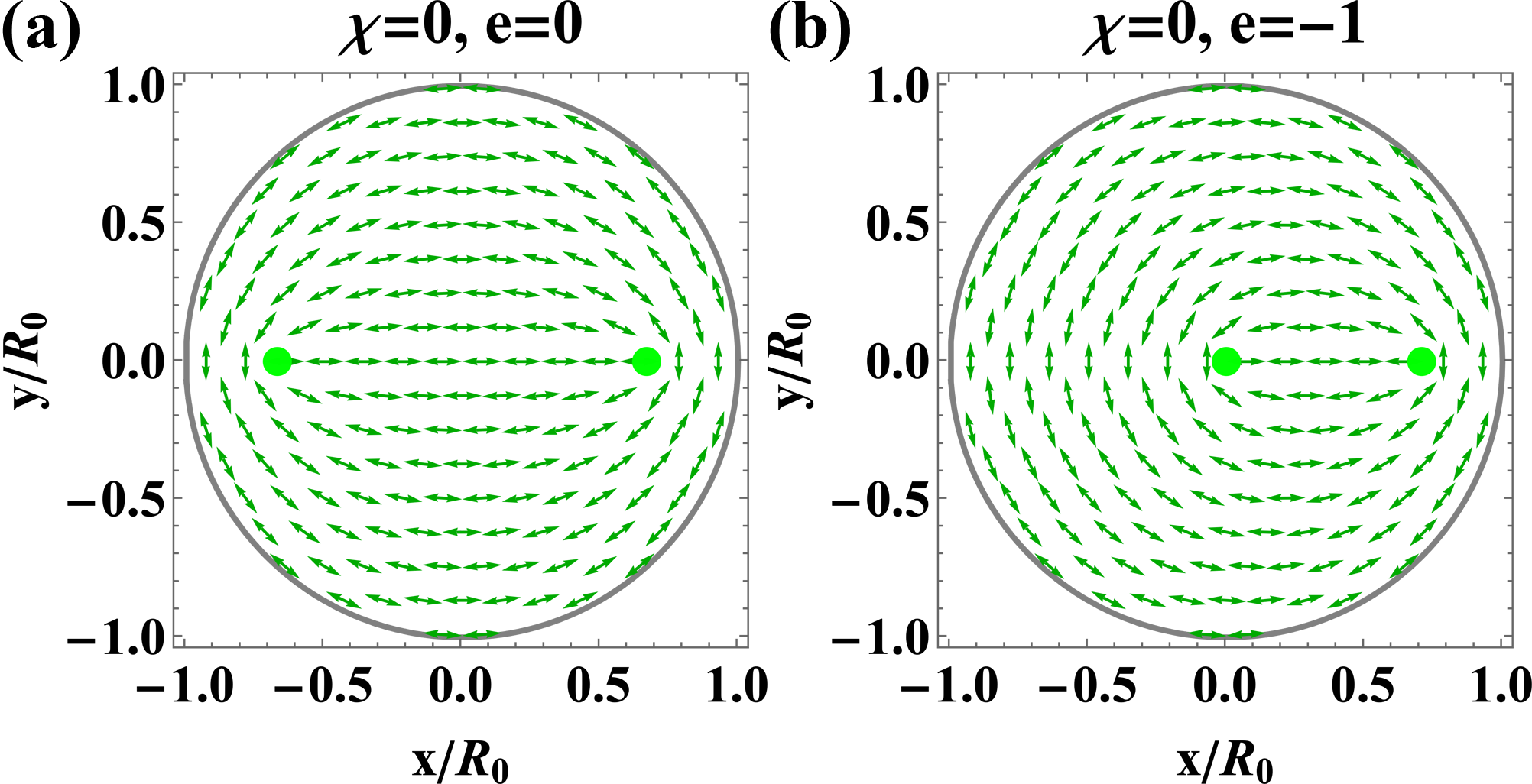}
\caption{\label{fig2p5:degenerateconfigs} Illustrations of (a) $e=0$ and (b) $e=-1$ configurations for $p=2$ liquid crystals on a 2D flat disk ($\chi=0$) of radius $R_0$ with tangential boundary conditions. Upon considering the lowest-order energy terms in our continuum theory, we find that the free energy difference between these two configurations is a constant independent of the system size $R_0$, and the configuration (a) is always stabler than (b), which is consistent with both Ref.~\cite{vafa2022defect} and earlier work in Ref.~\cite{duclos2017topological}. In the thermodynamic limit of $R_0/a \to \infty$, these two configurations are energetically close to each other because of the dominating self-energy terms.}
\end{figure*}
To understand the ground states in more detail, let's first consider the thermodynamic limit when the system size $R_0$ is much greater than the defect core size $a$ and $\bar{a} \to 0$. In this case, the last two terms in Eqn.~(\ref{eqn:totalfreeenergytangential}) dominate the free energy. By minimizing these dominant terms with respect to $e$, we find the total free energy has minimum at $e^* = -(1+(2p-1)\chi)/2$. However, $e$ is only allowed to take integer values, and one can see that as we increase $\chi$ continuously from $0$ to $1$, the actual value of $e$ minimizing the total free energy will make multiple jumps between consecutive integers, which means the total number of flank defects has discrete jumps as a function of $\chi$. To illustrate a discrete jump for the total number of flank defects in the thermodynamic limit, we inspect $\chi=0$ specifically. The ground state is a configuration where all the $+1/p$ defects are distributed symmetrically around the center of the disk, but since the self-energy of defects dominates over the interaction energy, the ground state is surrounded by many energetically similar configurations for any value of $p$: $e=0$ and $e=-1$ are energetically similar states, where the free energy difference between those two configurations is a finite constant, only weakly dependent on system size for the potential higher-order corrections in our free energy. If we look at $p=2$ liquid crystals on a 2D disk with $R_0 \gg a$ and tangential boundary conditions as an example (see Fig.~\ref{fig2p5:degenerateconfigs}), the near self-energy degeneracy at $\chi=0$ means that the configuration with two $+1/2$ defects distributed evenly around the center has the same divergent contribution to the self-energy as the configuration where one of two $+1/2$ defects is located at the center but the other one is off the center. In fact, we find that the former configuration has lower total free energy than the latter by a finite amount because of the interaction energy. As $\chi$ is increased slightly above $0$, the near self-energy degeneracy is broken due to the appearance of the pseudo-defect at the apex, and the advantage of $e=-1$ configuration in the dominating self-energy immediately wins over the finite advantage of $e=0$ configuration in the interaction energy. The only ground state becomes $e=-1$, corresponding to absorption of a single defect by the apex. More generally, the isolated values of $\chi$ at which the ground state has a two-fold self-energy degeneracy in $e$ are transition points where the total number of flank defects changes. We thus find that the transition between a configuration with $p-\gamma$ flank defects and $p-\gamma-1$ flank defects in the thermodynamic limit happens at
\begin{equation}
  \begin{aligned}
    \chi = \frac{2\gamma}{2p-1}.
  \end{aligned}
\end{equation}

\begin{figure*}[h]
\includegraphics[width=0.70\textwidth]{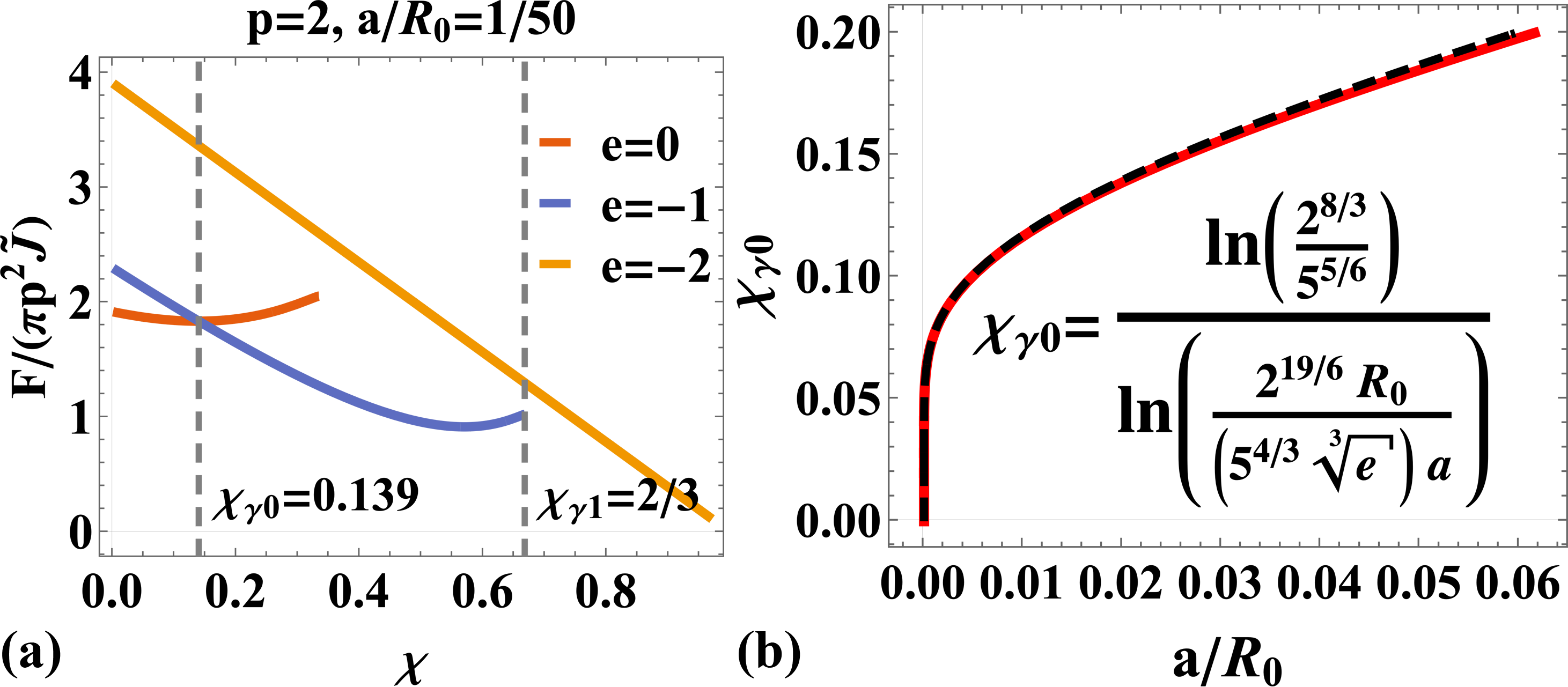}
\caption{\label{fig3:resultsfortangentialboundary} $p=2$ liquid crystal ground states on cones with tangential boundary conditions. (a) Free energy comparison among configurations with different number of $+1/2$ flank defects ($2+e$) at different $\chi$. (b) Transition point of $\chi_{\beta 0}$ between $e=0$ and $e=-1$ ground state as a function of $a/R_0$. The red solid line is solved numerically from our analytic theory, and the black dashed line is the asymptotic solution for $a/R_0 \to 0$. (The letter $e$ in the equation is the base of natural logarithm.)}
\end{figure*}
Our results thus far agree with the results of Ref.~\cite{vafa2022defect}. Next, using a specific example of $p=2$ liquid crystals, we show how the finite cone size shifts the transitions between ground states with different numbers of flank defects. When the system size $R_0$ is not infinite, in order to find the ground states, we have to consider all the terms in Eqn.~(\ref{eqn:totalfreeenergytangential}). Although there is no general analytical solution for minimizing the total free energy with respect to $e$ as a function of $\chi$ and $\bar{a}$, we can proceed numerically. In Fig.~\ref{fig3:resultsfortangentialboundary}(a), we display the total free energies for configurations with different numbers of flank defects as a function of $\chi$ assuming $\bar{a}=1/50$ and compare them to find how the ground state evolves as the shape of the conical surface changes. The free energy lines for $e=0$ and $e=-1$ can simply terminate when the flank defects for those two configurations are absorbed by the apex according to Eqn.~(\ref{eqn:defectpostangential}). Those end points should therefore fall onto the free energy line for $e=-2$, i.e. both flank defects absorbed by the apex. Due to nonuniversal differences between the short distance defect core radii for flank defects and the apex defect, the total free energies for $e=0$ and $e=-1$ are slightly lower than that for $e=-2$ at their end points as their flank defects are absorbed by the apex. 

For finite size cones, the conformal theory predicts three regimes in the $\chi$ axis. Unlike what we find in the thermodynamic limit, when $\chi<\chi_{\gamma 0}$, the ground state is always $e=0$. This difference can be easily understood in the conformal domain of a cone: As the cone size $R_0$ becomes finite, the conformal domain of a cone with small $\chi$ is a 2D disk of finite size $\tilde{R}_0$ containing two $+1/2$ defects, as well as a fixed pseudo-defect of charge $-\chi$ at the center. As we have discussed above, the $e=-1$ configuration has a lower self-energy compared to the $e=0$ configuration when $\chi>0$, which means the negative pseudo-defect at the apex still tends to absorb a $+1/2$ flank defect. The finite advantage of the $e=0$ configuration in the interaction energy, illustrated in Fig.~\ref{fig2p5:degenerateconfigs}(a), is now able to counter the defect absorption from the apex until $\chi=\chi_{\gamma 0}$. When $\chi_{\gamma 0}<\chi<\chi_{\gamma 1}$, the ground state changes into $e=-1$, where one of the two flank defects is absorbed to the apex. As $\chi$ passes through $\chi_{\gamma 1}$, the ground state becomes $e=-2$ as two flank defects are absorbed to the apex. To explore the dependence of the transition point $\chi_{\gamma 0}$ on the ratio of the defect core size to the cone size $\bar{a}=a/R_0$, we solve $\chi_{\gamma 0}$ as a function of $\bar{a}$ numerically in Fig.~\ref{fig3:resultsfortangentialboundary}(b), and find that the value of $\chi_{\gamma 0}$ increases monotonically as a function of $\bar{a}$. When $\bar{a}\to 0$, $\chi_{\gamma 0}$ tends to zero, agreeing with the result in the thermodynamic limit. For $R_0 \gg a$, we find that $\chi_{\gamma 0}$ is inversely proportional to $\ln{(R_0/a)}$ as $\bar{a}\to 0$. $\chi_{\gamma 1}$ itself does not depend on $\bar{a}$ because the end point of $e=-1$ cannot have higher free energy than $e=-2$, and remains locked at $\chi_{\gamma 1}=2/3$.

\section{Continuum Theory for Liquid Crystal Ground States on Cones with Anti-Twist Boundary Conditions}

Now, we use the total free energy in Eqn.~(\ref{eqn:totalfreeenergyndefect}) to investigate the ground states of $p$-atic liquid crystals on cones with anti-twist boundary conditions. When the system size $R_0$ is large compared to the defect core size $a$, the stablest defect in $p$-atic liquid crystals usually has topological charge of $\pm 1/p$. Hence, it is plausible to assume that the ground states are composed of a composite defect with charge $+e/p$ absorbed at the apex and $p+e$ defects of charge $-1/p$ on the flank. Here again, $e$ is an integer, and when $e$ is positive, $e$ represents the number of extra $-1/p$ defects on the flank, beyond those required by the total topological charge of $-1$ imposed by the anti-twist boundary conditions. Similar to the tangential boundary condition case~\cite{vafa2022defect}, we further assume that the $p+e$ defects on the flank are distributed evenly around the apex at $\tilde{z}_k = \tilde{r}_0 e^{i 2\pi k/(p+e)}$ where $k=1,2,\dotsc,p+e$, an assumption we confirm with the numerical simulations described in Section~\ref{sec:simulations}.

Upon inserting this defect ansatz into Eqn.~(\ref{eqn:totalfreeenergyndefect}), we get
\begin{equation}
  \begin{aligned}
    \frac{F}{\pi p^2 \tilde{J}} = -\frac{p+e}{p^2}\bigg\{ \ln{\left( (p+e) \bar{r}_0^{p+e-1} (1-\bar{r}_0^{2(p+e)}) \right)} + \\ 
    p\left( \frac{\chi}{2p} \!+\! \chi \!-\! \frac{e}{p} \right) \ln{\bar{r}_0^2}\! \bigg\} + \frac{p \!+\! e}{p^2} \ln{\frac{1 \!-\! \chi}{\bar{a}}} + \frac{(-\! \chi \!+\! e/p)^2}{1-\chi}\ln{\frac{1}{\bar{a}}},
  \end{aligned}
  \label{eqn:totalfreeenergyantitwist}
\end{equation}
where again, $\bar{r}_0=\tilde{r}_0/\tilde{R}_0$ is the dimensionless radial position of the flank defects in the conformal coordinates, and $\bar{a}=a/R_0$ is the ratio of the defect core size to the cone size in the fundamental domain. The simplified total free energy for anti-twist boundary conditions differs from Eqn.~(\ref{eqn:totalfreeenergytangential}) for tangential boundary conditions because the flank defects now carry negative charge instead of positive charge. With a fixed number of flank defects $p+e$ and fixed conical shape $\chi$, the free energy in Eqn.~(\ref{eqn:totalfreeenergyantitwist}) has minimum with respect to the position of flank defects at
\begin{equation}
  \bar{r}_0 = \left(\frac{p-e-1+\chi+2p\chi}{3p+e-1+\chi+2p\chi}\right)^{\frac{1}{2(p+e)}},
  \label{eqn:defectposantitwist}
\end{equation}
which is also slightly different from the result in Eqn.~(\ref{eqn:defectpostangential}) simply because there is repulsion between the apex pseudo-charge and the flank charges for anti-twist boundary conditions instead of attraction for tangential boundary conditions. For a flat disk with anti-twist boundary conditions, this result predicts that $p$ defects of charge $-1/p$ are distributed evenly around the center at $r_0/R_0=(\frac{p-1}{3p-1})^{1/2p}$, identical to the result for the positive defects associated with tangential boundary conditions. 

After inserting the optimized $\bar{r}_0$ into Eqn.~(\ref{eqn:totalfreeenergyantitwist}), we further minimize the total free energy to determine the optimal number of extra defects $e$ at fixed cone sharpness $\chi$ and cone size parameter $\bar{a}$. Similar to cones with tangential boundary conditions in the previous section, seeking a general analytic solution for $e$ as a function of $\chi$ and $\bar{a}$ is difficult. We first look at the thermodynamic limit of infinite cone size $\bar{a} \to 0$. The last two terms associated with $\bar{a}$ in Eqn.~(\ref{eqn:totalfreeenergyantitwist}) then dominate, and the minimum of the total free energy occurs at $e^*=((2p+1)\chi-1)/2$. Since $e$ only takes integer values, the integer closest to $e^*$ is the actual $e$ that the liquid crystals exploit to minimize the total free energy. The optimal value of $e$ thus changes discontinuously as a function of $\chi$. When $e^*$ is equidistant from two consecutive integers, the ground state is degenerate, implying that the transition between the configuration with $p+\gamma$ flank defects and that with $p+\gamma+1$ flank defects happens at
\begin{equation}
  \chi = \frac{2(\gamma+1)}{2p+1},
  \label{eqn:chitransitionlimit}
\end{equation}
where $\gamma=0,1,\dotsc,p-1$. Hence, the total number of flank defects increases as the cone becomes sharper, leaving more  $-1/p$ defects on the flank. However, because the total topological charge on the cone still has to be $-1$, there is additional positive topological charge at the apex in order to maximally cancel the $-\chi$ charge of the pseudo-defect. Remarkably, when $p=1$, our analytic theory predicts a transition between the configuration of $e=0$ and $e=1$ at $\chi=2/3$ in the thermodynamic limit of $\bar{a}\to 0$, a transition that has no analog on a cone with tangential boundary conditions.

\begin{figure*}[h!]
\includegraphics[width=0.70\textwidth]{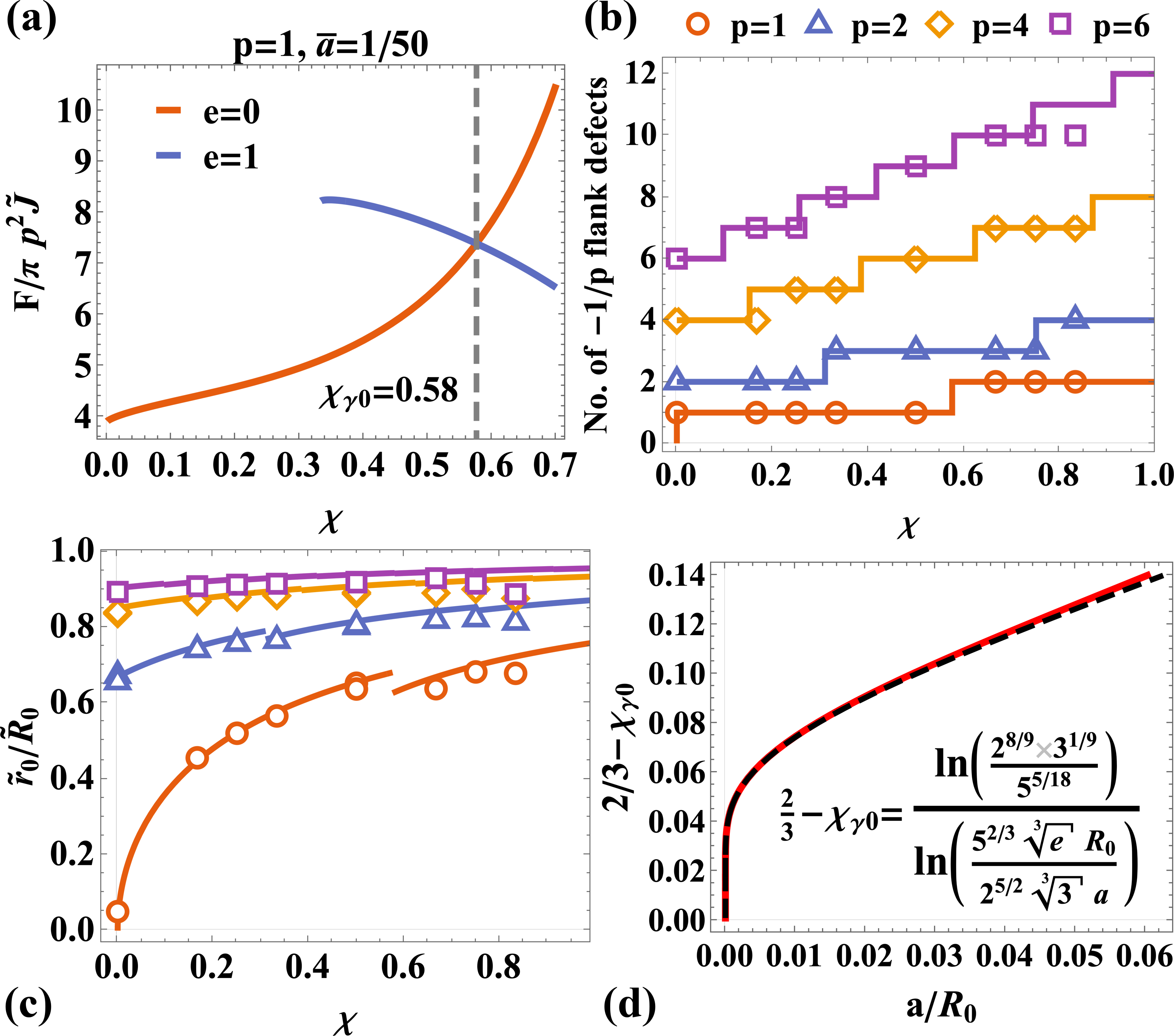}
\caption{\label{fig4:resultsforantitwistboundary} Analytic predictions and simulation results for liquid crystal ground states on cones with anti-twist boundary conditions. (a) Free energy comparison for $p=1$ liquid crystals between configurations with different number of $-1$ flank defects ($1+e$) at different $\chi$. The transition between $e=0$ and $e=1$ is denoted by $\chi_{\gamma 0}$. (b) Comparison between the number of $-1/p$ flank defects in the ground state from our theory (solid lines) and that from our simulations (data points) as a function of $\chi$. For both our analytic theory and simulation, $\bar{a}$ is fixed at $1/50$. (c) Comparison between the equilibrium radial position of the flank defects from our theory (solid lines) and that from our simulations (data points). The ordinate is rescaled by the size of the cone $\tilde{R}_0$. Here, $\bar{a}$ is also fixed at $1/50$. There are small jump discontinuities in $\tilde{r}_0/\tilde{R}_0$ at the transition values of $\chi$ indicated in panel (b). (d) Theoretical prediction for the transition $\chi_{\gamma 0}$ in $p=1$ liquid crystals as a function of $a/R_0$. The red solid line is solved numerically from our analytic theory, and the black dashed line is the asymptotic solution for $a/R_0 \to 0$. The quantity $e$ in the equation is the base of natural logarithm and does not represent the number of extra flank defects.}
\end{figure*}
When the ratio of the defect core size $a$ to the cone size $R_0$ is finite, we again have to consider all the terms in Eqn.~(\ref{eqn:totalfreeenergyantitwist}) to find the number of extra flank defects $e$ for the ground state at fixed $\chi$ and $\bar{a}$. We then calculate numerically the total free energies of configurations with different number of extra flank defects at fixed $\chi$ and $\bar{a}$ and compare them to find the lowest free energy state. Four types of liquid crystals with different rotational symmetries, $p=1,2,4,6$, are considered, and we set $\bar{a}=1/50$ for all of our numerical calculations. 

For the simplest case when $p=1$, the free energy comparison is shown in Fig.~\ref{fig4:resultsforantitwistboundary}(a). Similar to the thermodynamic limit $\bar{a} \to 0$, there are two classes of ground states for $p=1$ liquid crystals on cones with anti-twist boundary conditions, those with $e=0$ and $e=1$. When $\chi=0$, there is a single $-1$ defect located at the center of the flat disk. As $\chi$ becomes greater than zero, but not too large, the ground state retains a single $-1$ defect on the cone, but the $-1$ defect is repelled away from the apex because both the pseudo-defect charge at the apex and the actual defect charge have the same sign. After $\chi$ reaches a transition value $\chi_{\gamma 0}$, a pair of $\pm 1$ defects is created on the cone, leaving two $-1$ defects on the flank and one $+1$ defect at the apex. As we further increase the value of $\chi$, the total number of $-1$ defects remains $p+e=2$ while the position of the $-1$ defects is pushed closer to the boundary (due to the subtle fact that the free energy term associated with the varying defect core size in the conformal domain imposes a stronger repulsion to defects from the apex). The qualitative behavior of $p=1$ liquid crystals on anti-twist cones with $\bar{a}=1/50$ is the same as that we find in the limit of $\bar{a} \to 0$. Nevertheless, the finite system size in this case lowers the transition value from $\chi_{\gamma 0}=2/3$ to $\chi_{\gamma 0} \approx 0.58$.

In addition to $p=1$ liquid crystals, the free energy comparison can be done numerically for the other values of $p$. The solid lines in Fig.~\ref{fig4:resultsforantitwistboundary}(b) show how the total number of $-1/p$ flank defects in the ground state changes as a function of $\chi$. In contrast to liquid crystals on cones with tangential boundary conditions where the $+1/p$ defects are absorbed to the apex one after another as the cone becomes sharper~\cite{vafa2022defect}, for cones with anti-twist boundary conditions, we find that the intrinsic geometry of a conical surface triggers the release of more $-1/p$ defects to the flank in the ground state. This behavior follows from minimizing the charge of the pseudo-defect caused by the intrinsic geometry of a cone. Since the self-energy of the pseudo-defect at the apex is $\frac{\chi^2}{1-\chi}\ln{(R_0/a)}$, it is energetically preferable for the apex to absorb $+1/p$ defects to reduce the distortion free energy. Due to the anti-twist boundary conditions, the same number of $-1/p$ defects must appear on the flank to maintain the total topological charge on the cone of $-1$. 

The equilibrium position of the $-1/p$ flank defects in the ground states for different $p$-atic liquid crystals is displayed in Fig.~\ref{fig4:resultsforantitwistboundary}(c). This quantity is determined by a delicate interplay involving topological defects, boundary conditions, and the conical geometry. The flank defects are repelled closer to the cone base as the cone becomes sharper, mainly due to the smaller defect core size in the conformal domain. The discontinuities shown in the analytic lines for the equilibrium position of the flank defects $\tilde{r}_0/\tilde{R}_0$ are triggered by tiny first-order phase transitions between configurations with different number of flank defects.

Our analytic theory also explains how the ratio of the defect core size to the cone size $\bar{a}$ affects those transitions between configurations with different number of flank defects predicted in Eqn.~(\ref{eqn:chitransitionlimit}) in the thermodynamic limit $\bar{a} \to 0$. Our numerical calculations of the transition points as a function of $\bar{a}$ reveal that all the transition values are reduced compared to their thermodynamic limits for $p=1,2,4,6$, because the self-energy of flank defects is significantly lowered by the reduced area of the fundamental domain when the system size is finite. Fig.~\ref{fig4:resultsforantitwistboundary}(d) illustrates how the transition $\chi_{\gamma 0}$ between $e=0$ and $e=1$ configuration in $p=1$ liquid crystals evolves as a function of $\bar{a}$. By looking at the asymptotic behavior of the transitions in the limit of $\bar{a} \to 0$, we find a logarithmic correction to the transition between the configuration with $p+\gamma$ flank defects and that with $p+\gamma+1$ flank defects changes as a function of $\bar{a}$:
\begin{equation}
  \frac{2(\gamma+1)}{2p+1} - \chi_{\gamma} \sim \frac{1}{\ln{(R_0/a)}}.
\end{equation}

\section{Simulations for Liquid Crystals on Cones with Anti-Twist Boundary Conditions} \label{sec:simulations}

\begin{figure*}[h]
\includegraphics[width=0.99\textwidth]{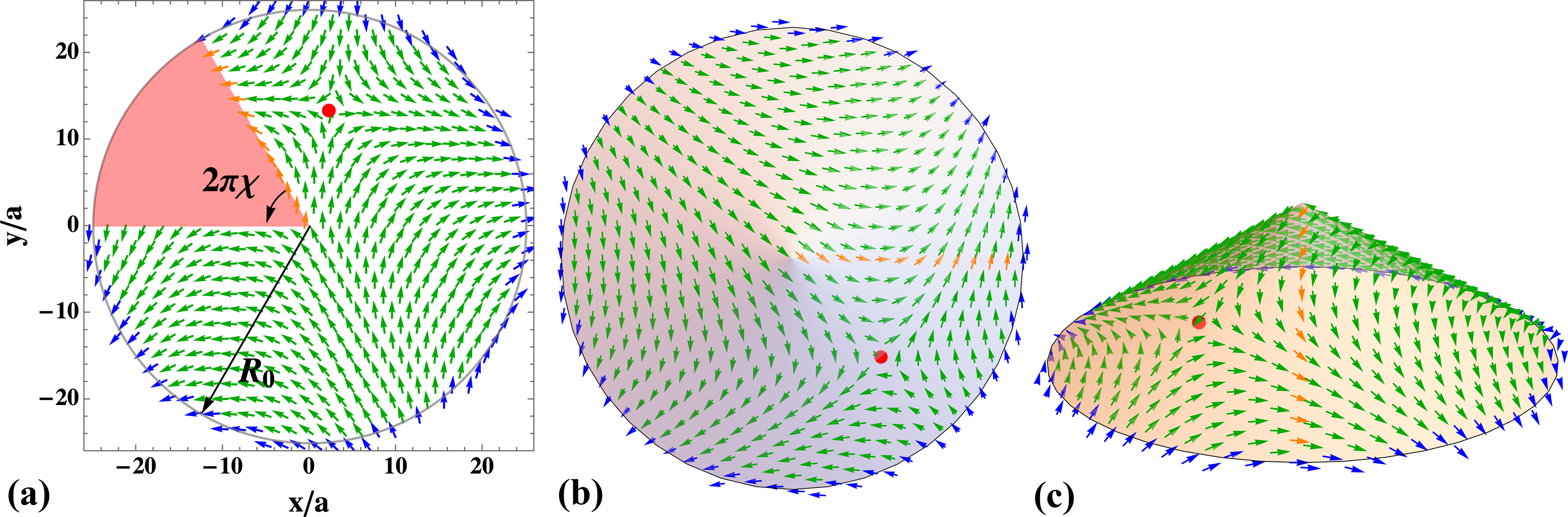}
\caption{\label{fig5:simulationmodel} Illustration of our simulation model. In panel (a), the order parameter field of a $p$-atic liquid crystal with $p=1$ on a cone of sharpness $\chi$ is simulated on a triangular lattice with lattice constant $a$ in a circular region of radius $R_0$ by removing a slice of the disk (red region). Here, $\chi=1/6$, corresponding to a deficit angle $2\pi \chi=\pi/3$. The green arrows inside the disk represent the orientation of the order parameter field and have an aligning interaction between nearest neighbors. The blue arrows close to the edge of the disk are fixed, implementing the anti-twist boundary conditions. The orange arrows on one side of the removed region are the duplicates of the green arrows on the other side of the red region rotated by the parallel transport angle $-2\pi \chi$, satisfying the continuity condition in Eqn.~(\ref{eqn:continuityfundamentaldomain}). The $-1$ flank defect predicted for anti-twist boundary conditions when $\chi=1/6$ is shown as the red dot. The actual radius of the disk in this illustration is $25$, but the density of the visualized order parameter field is lowered on purpose for better visualization. (b)-(c) Same configuration as in (a), but visualized on a cone of $\chi=1/6$ in three dimensions. (b) the top view; (c) the front view.}
\end{figure*}
To test our theoretical predictions, the order parameter field $\boldsymbol{u}$ of $p$-atic liquid crystals on a cone of radius $R_0$ and sharpness $\chi$ can be conveniently modeled on either a triangular lattice or a square lattice at commensurate values of $\chi$ ($\chi=0$, $1/6$, $1/4$, $2/6$, $3/6=2/4$, $4/6$, $3/4$, and $5/6$) on an unrolled cone where the lattice is trimmed to the shape of the fundamental domain~\cite{zhang2022fractional}. Our simulation model with anti-twist boundary conditions is illustrated for $p=1$ in Fig.~\ref{fig5:simulationmodel}(a). We start from a triangular or a square lattice in a 2D flat space where the origin of the coordinate system is assigned to an arbitrary lattice site, and the $x$-direction is aligned with one of the two nearest neighbor lattice vectors. The lattice constant $a$ is fixed at $1$ in our simulations, and we choose $R_0=50$. The continuous order parameter field $\boldsymbol{u}$ on a cone is then discretized by the green arrows on the lattice sites inside a disk of radius $R_0$ centered at the origin. The lattice sites in an area between the polar angle $\pi - 2\pi \chi$ and $\pi$ (red region in Fig.~\ref{fig5:simulationmodel}(a)) are removed so that the trimmed lattice has the shape of an unrolled cone cut along one of its flank geodesics. Here, angles are measured from the positive $x$-axis. The orientation of $\boldsymbol{u}_{i}$  at each lattice site can simply be described by an angle $\psi_{i}$ via the relation $\boldsymbol{u}_{i}=\hat{\boldsymbol{x}}\cos{\psi_{i}}+\hat{\boldsymbol{y}}\sin{\psi_{i}}$. 

The Hamiltonian of the discretized field for $p$-atics reads
\begin{equation}
  H = J \sum_{\langle ij \rangle}{\left( 1-\cos{\left(p(\psi_i -\psi_j)\right)}\right)}.
  \label{eqn:simulationhamiltonian}
\end{equation}
Eqn.~(\ref{eqn:fundamentaldomain}) can be regarded as the continuum version of this Hamiltonian, upon taking account of the lowest order terms in $\psi_i - \psi_j$ on the right-hand side. $J$ is the interaction strength between the order parameter $\psi_i$ and its nearest neighbors, and we choose energy units such that $J=1$ for both our simulations in triangular lattices and square lattices. Although $J$ in our simulations is chosen to be the same, the effective interaction strength $\tilde{J}$ in our continuum theory depends on both $J$ and the lattice type. For square lattices, $\tilde{J}=J$; for triangular lattices, $\tilde{J}=\sqrt{3}J$~\cite{zhang2022fractional}. Because we only consider nearest-neighbor interactions, the summation $\langle ij \rangle$ runs over all pairs of nearest neighbors in the fundamental domain. The integer $p$ describes the rotational symmetry of the order parameter.

To implement the boundary conditions at the base of a cone, we freeze the directions of the blue arrows in Fig.~\ref{fig5:simulationmodel}(a) outside the disk of radius $R_0$, which then interact with the green arrows next to them. Since the anti-twist boundary conditions are considered, the orientation of the blue arrows is determined by
\begin{equation}
  \psi_i = -\frac{1+\chi}{1-\chi} (\phi_i + \pi) -\frac{\pi}{2},
\end{equation}
so that the orientation of the order parameter at the boundary is rotated by $-2\pi-2\pi \chi$ moving from $\phi=-\pi$ to $\phi=\pi-2\pi \chi$. Because the simulations are carried out in the fundamental domain, the continuity condition in Eqn.~(\ref{eqn:continuityfundamentaldomain}) has to be satisfied as well at the two cut edges of the removed region. In order to apply this continuity condition in our simulations, we need to smoothly join the lattice sites at the two edges. Accordingly, the discrete rotational symmetry of triangular lattices or square lattices only allows for certain values of $\chi$. For triangular lattices, the available values of $\chi$ are $0$, $1/6$, $2/6$, $3/6$, $4/6$, and $5/6$; for square lattices, the available values are $0$, $1/4$, $2/4$, and $3/4$. Due to the continuity condition at the two cut edges, those orange arrows in Fig.~\ref{fig5:simulationmodel}(a) at $\phi=\pi-2\pi \chi$ are duplicates of the order parameter at the lattice sites of $\phi=-\pi$ plus a rotation by the parallel transport angle $-2\pi \chi$. The validity of the continuity condition is further demonstrated by visualizing the discretized order parameter field on a 3D conical surface in Fig.~\ref{fig5:simulationmodel}(b-c). Because the order parameter at the apex is not well-defined, we remove the lattice site at the origin so that the order parameter at the origin does not interact with its nearest neighbors, which technically creates a small hole of radius $a$ at the origin with free boundary conditions in our simulations. The removed apex only has a minor effect on liquid crystal ground states on cones, provided its size is small enough compared to the cone dimensions~\cite{vafa2022defect}. It is the continuity condition at the two cut edges that imposes the conical geometry. This inner free boundary condition for the removed apex slightly changes the core energy of topological defects located at the apex. To minimize the difference between our simulation model and our analytic calculations, we have defined a larger defect core size for topological defects at the apex in our analytic theory.

\begin{figure*}
\includegraphics[width=0.99\textwidth]{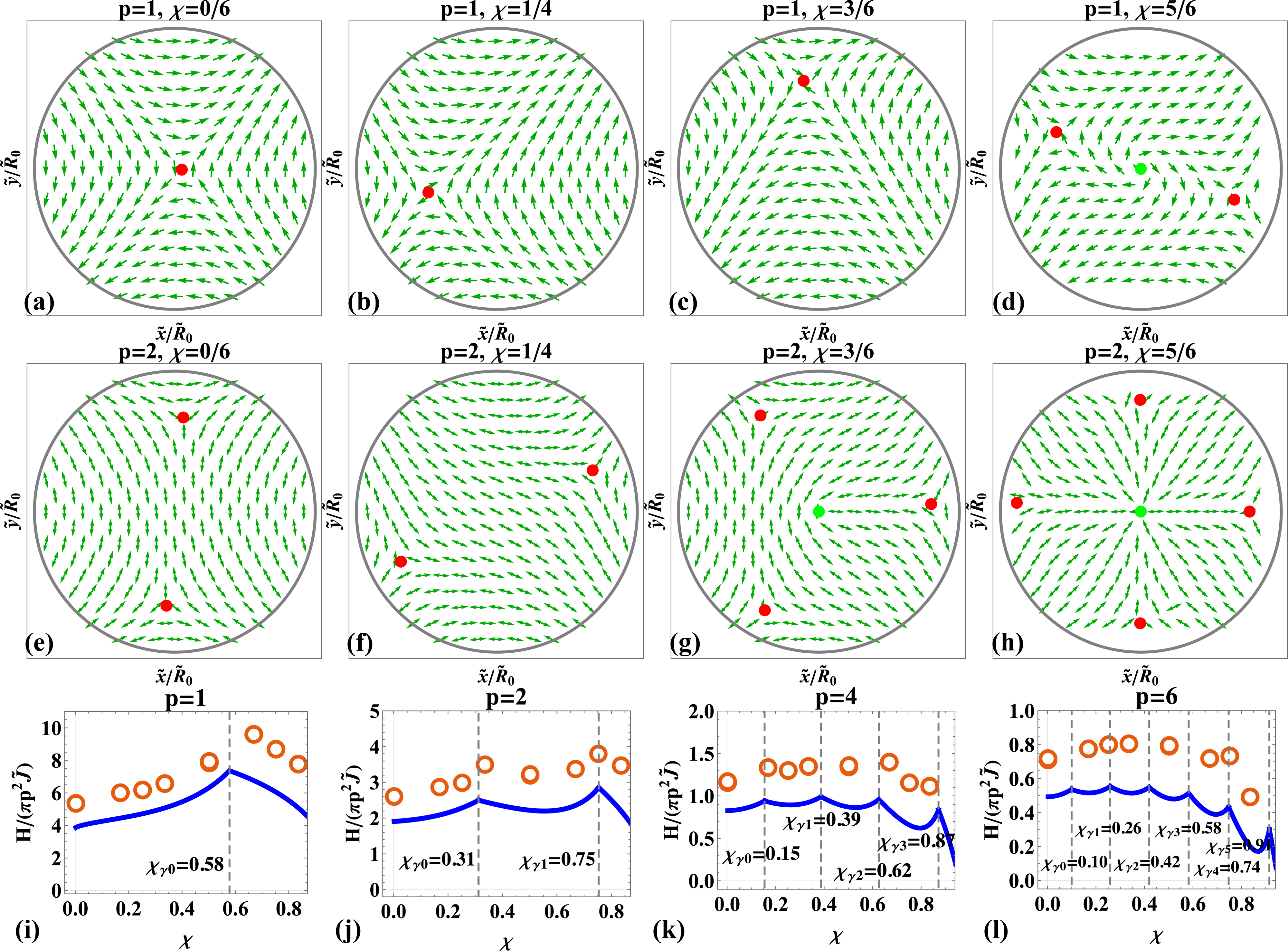}
\caption{\label{fig:simulationresults}Simulation results. (a)-(d) $p=1$ liquid crystal ground states visualized by the order parameter field $\psi_i' = \psi_i + \chi \tilde{\phi}_i$ in the conformal domain at different values of $\chi$ with anti-twist boundary conditions. The green arrows represents the orientation of the polar order parameter $\boldsymbol{u}_i' = \hat{\boldsymbol{x}}\cos{\psi_i'} + \hat{\boldsymbol{y}}\sin{\psi_i'}$ on the cone. The red dots are $-1$ defects, and the green dots are $+1$ defects. (e)-(h) $p=2$ liquid crystal ground states visualized by the order parameter field $\psi_i'$ in the conformal domain at different values of $\chi$ with anti-twist boundary conditions. The green double-headed arrows are the orientations of the local nematic director on the cone. The red dots are $-1/2$ defects, and the green dots are plus defects whose charge depends on $\chi$. (i)-(l) Comparison of the ground state energy calculated from our simulations (orange circles) and from the analytic theory (solid blue lines).}
\end{figure*}
We search for the ground states of $p$-atic liquid crystals on cones with the discrete lattice model by slowly annealing the order parameter field from high temperature $k_B T/J = 1.2$ to zero temperature $k_B T/J=0$ with a Langevin thermostat~\cite{mccarthy1986numerical,bray1994theory,berthier2001nonequilibrium}. To accurately find the ground state at each available value of $\chi$, we perform the same simulation ten times with different random seeds and use the configuration with the lowest free energy among them to approximate the ground state. The simulation results for $p=1$ and $p=2$ liquid crystals are visualized in the conformal domain in Fig.~\ref{fig:simulationresults}(a)-(h). Even though the lattice spacing becomes inhomogeneous in the conformal domain, for better visualization, we interpolate the order parameter field onto a uniform triangular lattice. When $p=1$, as expected, the ground state for $\chi=0$ only includes one $-1$ defect located at the center of the flat disk. As $\chi$ is increased, the only $-1$ defect is pushed off the apex due to the emergent negative pseudo-defect, which breaks the rotational symmetry. Beyond a threshold value of $\chi$, an extra $-1$ defect is created in the ground state, accompanied by a $+1$ defect located at the apex. We find that those two $-1$ defects are distributed symmetrically around the apex, consistent with the assumptions of our analytic theory. For $p=2$, the behavior of the ground state as a function of $\chi$ is qualitatively similar to that at $p=1$. The ground state starts from two $-1/2$ defects distributed symmetrically around the center at $\chi=0$. The positions of the two $-1/2$ defects are pushed closer to the boundary as $\chi$ increases, until extra $\pm 1/2$ defect pairs are created when $\chi$ exceeds various transition values. The positive defects stay at the apex, while the negative defects go to the flank. The evolution of the ground states for $p=4$ and $p=6$ as a function of $\chi$ shares the same pattern as $p=1$ and $p=2$, but for larger $p$, there are more $-1/p$ defects created on the flank at large $\chi$ in order to maximally cancel the implicit $-\chi$ pseudo-defect at the apex. 

Ground state configurations can be quantified by the number of the $-1/p$ defects on the flank and the positions of the flank defects. Ground states derived from our simulations for $p=1,2,4,6$ are summarized in Fig.~\ref{fig4:resultsforantitwistboundary}(b) and (c), and compared with our analytic results. Our simulation results agree well with the analytic predictions except that the positions of the flank defects at the largest $\chi$ in our simulations are slightly smaller than the theoretical values. We believe this discrepancy arises because at large $\chi$, the area of the fundamental domain is so small that the effect of the lattice spacing becomes significant at the resolution of our simulations.

We also calculate the total distortion free energy in Eqn.~(\ref{eqn:simulationhamiltonian}) from our simulations and compare it with our analytic theory in Fig.~\ref{fig:simulationresults}(i)-(l). Our analytic theory only considers the lowest order gradient expansion term from Eqn.~(\ref{eqn:simulationhamiltonian}). We believe this truncation accounts for the constant offset between the distortion free energy from our simulation and that from the theory. If the constant offset due to short distance physics is subtracted off, the trend of the distortion free energy from our simulation data fits well with our theoretical prediction. The local maxima caused by the micro-first-order transitions between configurations with different number of flank defects seem well reflected in our simulation results. 

\section{Conclusions}

In this work, we have developed a simple and intuitive theoretical framework for understanding the intrinsic effects of the concentrated Gaussian curvature at the apex of a conical surface coated with $p$-atic liquid crystals lying in a tangent plane, building on the work in Ref.~\cite{zhang2022fractional,vafa2022defect}. With the help of the conformal mapping, the irregular shape of the fundamental domain of an unrolled cone is transformed to a full disk in the conformal domain, and the free energy density of $p$-atic liquid crystals in the conformal domain remains the same form as that of 2D liquid crystals in a flat surface, except that the concentrated Gaussian curvature at the apex acts as a fixed unquantized pseudo-defect of charge $-\chi$ at the center of the conformal domain~\cite{vafa2022defect}. This $-\chi$ pseudo-defect at the apex defines how the parallel transport on a conical surface affects the order parameter field of liquid crystals and interacts with the topological defects.

We have analyzed $p$-atic liquid crystals on cones with free boundary conditions, tangential boundary conditions, and anti-twist boundary conditions, thus making contact with prior work for free and tangential boundary conditions~\cite{zhang2022fractional,vafa2022defect}. Our results for liquid crystals on cones with anti-twist boundary conditions are new. For free boundary conditions where there is no constraint on the total topological charge on the cone, liquid crystal ground state configurations on the cone absorb quantized positive topological defects to the apex in order to maximally cancel the $-\chi$ pseudo-defect as $\chi$ increases, consistent with the previous study in Ref.~\cite{zhang2022fractional}. For tangential boundary conditions, we confirm that the $+1/p$ defects created by these boundary conditions on the flank of the cone are gradually absorbed by the apex at specific transition points of $\chi$ as the cone becomes sharper~\cite{vafa2022defect}. Our analytic theory reveals that the transitions between ground states with different number of flank defects have a logarithmic dependence on the size of the cone and that the transitions predicted in Ref.~\cite{vafa2022defect} correspond to a cone of $R_0 \to \infty$.

To investigate how the concentrated positive Gaussian curvature at the apex of a cone interacts with negatively charged topological defects, we have performed both analytic theory and simulations studying liquid crystal ground states on cones with anti-twist boundary conditions at the cone base, which constrains the total topological defect charge to be $-1$. Both our analytic theory and simulations reveal that, as the cone becomes sharper with increasing $\chi$, the ground state for $p$-atic liquid crystals contains more $-1/p$ defects on cone flanks than the number of $-1/p$ defects for a disk geometry with $\chi=0$. The extra $-1/p$ defects are compensated by an equal number of $+1/p$ defects absorbed at the apex, maintaining the total topological charge on the cone at $-1$. The extra positive defects at the apex partially cancel the increasing distortion free energy from the $-\chi$ pseudo-defect. Unlike the case of tangential boundary conditions where a single $+1$ defect keeps staying at the apex no matter how sharp the cone is, the $p=1$ texture with anti-twist boundary conditions breaks rotational symmetry as soon as $\chi>0$. As the ratio of the defect core size $a$ to the cone size $R_0$ increases from zero, the critical values of $\chi_{\gamma}$ shift from their values in the thermodynamic limit by amounts of order $1/\ln{(R_0/a)}$.

\begin{acknowledgments}
We are indebted to Farzan Vafa for many helpful conversations, for a careful reading of the manuscript and for encouraging us to look more carefully at the contributions to the defect self-energy in both Fig.~\ref{fig2p5:degenerateconfigs} and Appendix~\ref{appendixa}. We also acknowledge helpful conversations with Grace Zhang. This research was supported in part by the National Science Foundation, through the Harvard University Materials Research Science and Engineering Center under grant DMR-2011754.
\end{acknowledgments}

\appendix

\section{Simulation demonstration for the self-energy of flank defects}
\label{appendixa}

In Section \MakeUppercase{\romannumeral 2}, we have derived the total free energy of a $N$-defect configuration on a cone with either anti-twist or tangential boundary conditions at the cone base in Eqn.~(\ref{eqn:totalfreeenergyndefect}), where the self-energy of a flank defect with topological charge $\sigma_k$ on the cone is identified to be
\begin{equation}
  \sigma_k^2 \ln{\frac{(1-\chi)R_0}{a}}
\end{equation}
dependent on $\chi$. In this appendix, with the help of lattice simulations, we briefly demonstrate that the $1-\chi$ factor inside the logarithm is detectable in the total free energy for the finite-size cones accessible in our simulations.

Since the $1-\chi$ factor is inside a logarithm, we can separate the $1-\chi$ factor from $R_0/a$ and break the self-energy into two terms. One of them is $\sigma_k^2 \ln{(1-\chi)}$, and the other is $\sigma_k^2 \ln{(R_0/a)}$. After that, we reorganize the total free energy in Eqn.~(\ref{eqn:totalfreeenergyndefect}) to
\begin{equation}
  \frac{F}{\pi p^2 \tilde{J}} = \alpha \ln{\frac{R_0}{a}} + F_{res}.
\end{equation}
In the first term on the right hand side of the equation above, all terms proportional to $\ln{(R_0/a)}$ in Eqn.~(\ref{eqn:totalfreeenergyndefect}) are grouped together such that
\begin{equation}
  \alpha = \sum_{k=1}^N{\sigma_k^2} + \frac{(\sigma_0 - \chi)^2}{1-\chi}.
\end{equation}
All the remaining terms in Eqn.~(\ref{eqn:totalfreeenergyndefect}) are included in $F_{res}$ which mainly involves the interactions between defects, as well as the separated $\ln{(1-\chi)}$ terms. As illustrated in our simulations, the defects are distributed symmetrically around the apex in the ground states. With this assumption, we infer the ratio of the defect radial position to the flank radius is independent of $R_0/a$ in the liquid crystal ground states on cones with either tangential or anti-twist boundary conditions, shown in Eqn.~(\ref{eqn:defectpostangential}) and Eqn.~(\ref{eqn:defectposantitwist}). Thus, $F_{res}$ is actually independent of $R_0/a$ for ground state configurations with both tangential and anti-twist boundary conditions but still dependent on $\chi$.

\begin{figure*}
\includegraphics[width=0.99\textwidth]{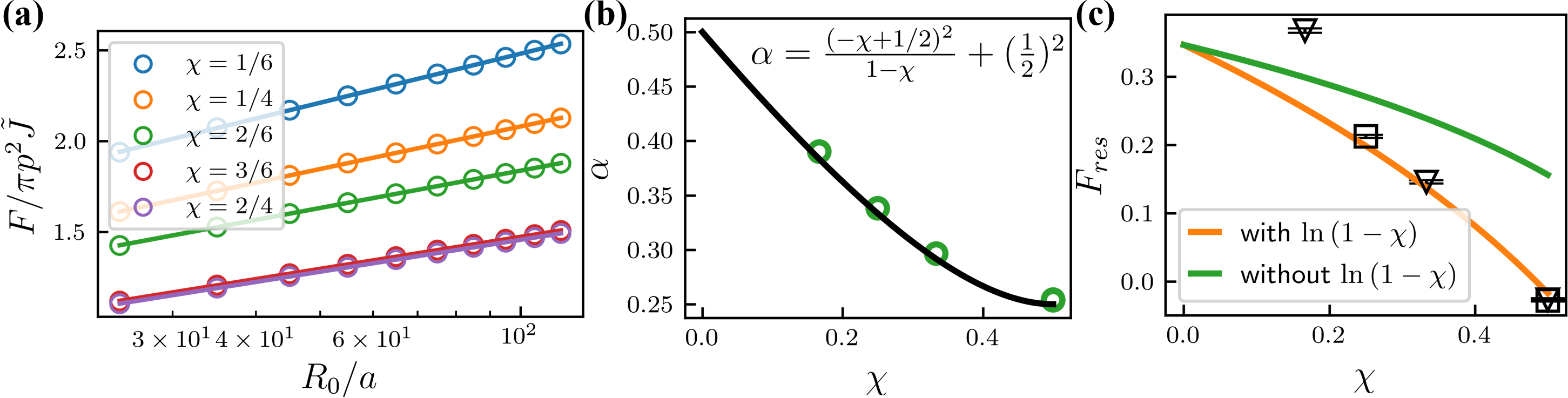}
\caption{\label{fig:appendix1}(a) Normalized ground state free energy of $p=2$ liquid crystal on cones as a function of the cone size $R_0/a$, calculated from the lattice simulations. A logarithmic scale is used on the horizontal axis. Circles with the same color represent the simulation data from the same value of $\chi$. The colored solid lines are fitted lines to the corresponding data sets. (b) The coefficient $\alpha$ vs the cone sharpness $\chi$. The green circles are the extracted values of $\alpha$ from data sets in (a). The black solid line is our theoretical prediction. (c) $F_{res}$ vs $\chi$. The black triangles (squares) mark the mean values of $F_{res}$ calculated from different cone size $R_0$ in triangular lattices (square lattices), and the black horizontal lines inside the data points are error bars. The orange solid line is from our analytic prediction with the contribution of $\ln{(1-\chi)}$ terms, and the green solid line is from the prediction ignoring the $\ln{(1-\chi)}$ terms.}
\end{figure*}

To detect the presence of the $\ln{(1-\chi)}$ terms obtained from the self-energy of flank defects on finite cones in $F_{res}$, we extract the coefficient $\alpha$ of the $\ln{(R_0/a)}$ term from the total free energies calculated from our lattice simulations in different flank radii $R_0$, retrieve $F_{res}$ from our simulation data by subtracting $\alpha \ln{(R_0/a)}$ from the total free energy, and compare the resulting $F_{res}$ as a function of $\chi$ with our theoretical predictions for $F_{res}$ with and without the $\ln{(1-\chi)}$ contribution. Our simulation results are summarized in Fig.~\ref{fig:appendix1}.

Liquid crystals with $p=2$ are simulated on cones with tangential boundary conditions at the cone base, with methods similar to what we have applied to anti-twist boundary conditions in Section \MakeUppercase{\romannumeral 4}. In these simulations, the lattice constant $a$ is fixed at $1$; the energy unit is chosen properly so that $J=1$. To search for the ground state at fixed $\chi$ and $R_0$, we slowly annealed a random initial configuration from a high temperature to zero temperature multiple times with different random seeds and pick the lowest energy state among them as the ground state. The calculated free energy of $p=2$ liquid crystal ground states is plotted as a function of the cone size $R_0$ in Fig.~\ref{fig:appendix1}(a) for different $\chi$. The cone size $R_0$ ranges from $25$ to $115$ at an interval of $10$. We choose $\chi=1/6$, $1/4$, $2/6$, and $3/6$ ($2/4$) so that the $p=2$ ground states at those values of $\chi$ have the same number of flank defects (one $+1/2$ flank defect while the other $+1/2$ defect is absorbed by the apex). For cones of $\chi=1/2$, simulations are performed both for triangular lattices ($\chi=3/6$) and for square lattices ($\chi=2/4$). As shown in Fig.~\ref{fig:appendix1}(a), the free energies for $\chi=3/6$ at different cone size almost overlap with those for $\chi=2/4$ except that the free energies of $\chi=2/4$ are slightly lower. We believe this difference arises because the free energy of a defect calculated in square lattices has a smaller core energy compared to triangular lattices with the same lattice constant. With the ground state free energy calculated from our simulation data, the corresponding $\alpha$ is extracted in Fig.~\ref{fig:appendix1}(b) for different $\chi$, compared with the theoretical prediction. One can see that our analytic prediction for $\alpha$ agrees very well with our simulation data. We can now subtract $\alpha \ln{(R_0/a)}$ from those free energies to derive $F_{res}$ for each $R_0$ and $\chi$. We expect that $F_{res}$ determined in this way will be independent of $R_0$. In Fig.~\ref{fig:appendix1}(c), we plot the mean value of $F_{res}$ derived from the different cone size $R_0$ as a function of $\chi$, together with the corresponding standard deviation as the error bars. As expected, the standard deviations are small for all $\chi$. To reconcile the difference in core energies for square and triangular lattices in Fig.~\ref{fig:appendix1}(c), we vertically shifted the mean values of $F_{res}$ so that the mean value of $F_{res}$ for $\chi=3/6$ overlaps with that for $\chi=2/4$. The trend of $F_{res}$ as a function of $\chi$ from our simulation data shows better agreement with the theoretical prediction including the $\ln{(1-\chi)}$ terms for the system sizes we have considered. The deviation at $\chi=1/6$ may be due to nearby metastable states in our finite systems.

\nocite{*}

\bibliography{apssamp}

\end{document}